\documentclass[twoside,twocolumn,9pt]{article}
\usepackage{extsizes}
\usepackage[super,sort&compress,comma]{natbib} 
\usepackage[version=3]{mhchem}
\usepackage[left=1.5cm, right=1.5cm, top=1.785cm, bottom=2.0cm]{geometry}
\usepackage{balance}
\usepackage{times,mathptmx}
\usepackage{sectsty}
\usepackage{graphicx,epsfig,overpic,hyperref}
\usepackage[normalem]{ulem}
\usepackage{graphicx} 
\usepackage{lastpage}
\usepackage[format=plain,justification=justified,singlelinecheck=false,font={stretch=1.125,small,sf},labelfont=bf,labelsep=space]{caption}
\usepackage{float}
\usepackage{fancyhdr}
\usepackage{fnpos}
\usepackage[english]{babel}
\addto{\captionsenglish}{%
  
}
\usepackage{array}
\usepackage{droidsans}
\usepackage{charter}
\usepackage[T1]{fontenc}
\usepackage[usenames,dvipsnames]{xcolor}
\usepackage{setspace}
\usepackage[compact]{titlesec}
\usepackage{hyperref}
\usepackage{amsmath,amsfonts,bm, amssymb}
\usepackage{graphicx,epsfig,overpic,hyperref}
\usepackage{xcolor}
\def\beq{\begin{equation}}
\def\eeq{\end{equation}}
\def\bea{\begin{eqnarray}} 
\def\eea{\end{eqnarray}}
\newcommand{\tp}[1]{#1}

\usepackage{epstopdf}

\definecolor{cream}{RGB}{222,217,201}

\begin{document}

\pagestyle{fancy}
\thispagestyle{plain}
\fancypagestyle{plain}{

\renewcommand{\headrulewidth}{0pt}
}

\makeFNbottom
\makeatletter
\renewcommand\LARGE{\@setfontsize\LARGE{15pt}{17}}
\renewcommand\Large{\@setfontsize\Large{12pt}{14}}
\renewcommand\large{\@setfontsize\large{10pt}{12}}
\renewcommand\footnotesize{\@setfontsize\footnotesize{7pt}{10}}
\makeatother

\renewcommand{\thefootnote}{\fnsymbol{footnote}}
\renewcommand\footnoterule{\vspace*{1pt}%
\color{cream}\hrule width 3.5in height 0.4pt \color{black}\vspace*{5pt}} 
\setcounter{secnumdepth}{5}

\makeatletter 
\renewcommand\@biblabel[1]{#1}            
\renewcommand\@makefntext[1]%
{\noindent\makebox[0pt][r]{\@thefnmark\,}#1}
\makeatother 
\renewcommand{\figurename}{\small{Fig.}~}
\sectionfont{\sffamily\Large}
\subsectionfont{\normalsize}
\subsubsectionfont{\bf}
\setstretch{1.125} 
\setlength{\skip\footins}{0.8cm}
\setlength{\footnotesep}{0.25cm}
\setlength{\jot}{10pt}
\titlespacing*{\section}{0pt}{4pt}{4pt}
\titlespacing*{\subsection}{0pt}{15pt}{1pt}

\fancyfoot{}
\fancyfoot[LO,RE]{\vspace{-7.1pt}\includegraphics[height=9pt]{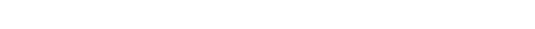}}
\fancyfoot[CO]{\vspace{-7.1pt}\hspace{13.2cm}\includegraphics{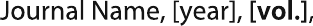}}
\fancyfoot[CE]{\vspace{-7.2pt}\hspace{-14.2cm}\includegraphics{RF}}
\fancyfoot[RO]{\footnotesize{\sffamily{1--\pageref{LastPage} ~\textbar  \hspace{2pt}\thepage}}}
\fancyfoot[LE]{\footnotesize{\sffamily{\thepage~\textbar\hspace{3.45cm} 1--\pageref{LastPage}}}}
\fancyhead{}
\renewcommand{\headrulewidth}{0pt} 
\renewcommand{\footrulewidth}{0pt}
\setlength{\arrayrulewidth}{1pt}
\setlength{\columnsep}{6.5mm}
\setlength\bibsep{1pt}

\makeatletter 
\newlength{\figrulesep} 
\setlength{\figrulesep}{0.5\textfloatsep} 

\newcommand{\topfigrule}{\vspace*{-1pt}%
\noindent{\color{cream}\rule[-\figrulesep]{\columnwidth}{1.5pt}} }

\newcommand{\botfigrule}{\vspace*{-2pt}%
\noindent{\color{cream}\rule[\figrulesep]{\columnwidth}{1.5pt}} }

\newcommand{\dblfigrule}{\vspace*{-1pt}%
\noindent{\color{cream}\rule[-\figrulesep]{\textwidth}{1.5pt}} }

\makeatother

\twocolumn[
  \begin{@twocolumnfalse}
  {\includegraphics[height=30pt]{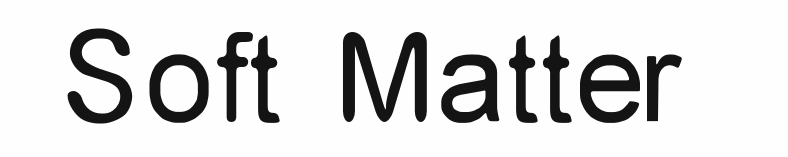}\hfill%
 \raisebox{0pt}[0pt][0pt]{\includegraphics[height=55pt]{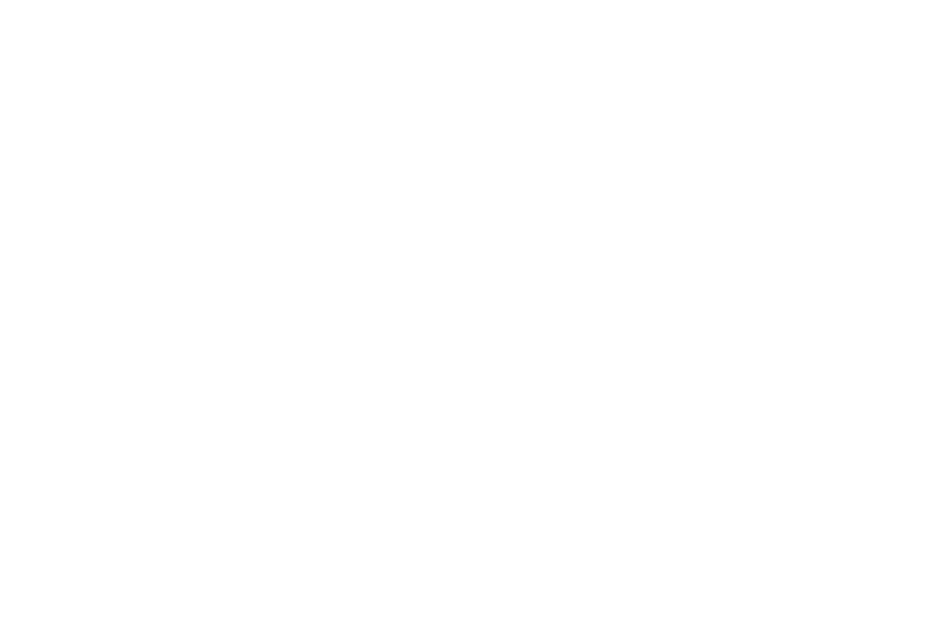}}%
 \\[1ex]%
 \includegraphics[width=18.5cm]{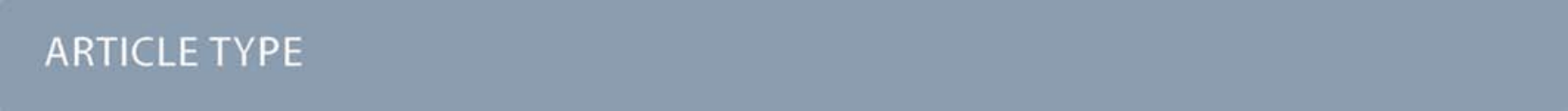}}\par
\vspace{1em}
\sffamily
\begin{tabular}{m{4.5cm} p{13.5cm} }

\includegraphics{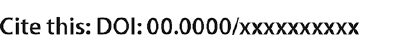} & \noindent\LARGE{\textbf{Axisymmetric membranes with edges under external force: buckling, minimal surfaces, and tethers}}$^\dag$ \\
\vspace{0.3cm} & \vspace{0.3cm} \\

 & \noindent\large{Leroy L.  Jia,$^{\ast}$\textit{$^{a}$} Steven Pei,\textit{$^{b}$} Robert A. Pelcovits\textit{$^{b}$} and  Thomas R. Powers \textit{$^{cb}$}} \\

\includegraphics{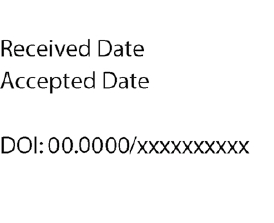} & \noindent\normalsize{We use theory and numerical computation to determine the shape of an axisymmetric fluid membrane with a resistance to bending and constant area. The membrane connects two rings in the classic geometry that produces a catenoidal shape in a soap film. In our problem, we find infinitely many branches of solutions for the shape and external force as functions of the separation of the rings, analogous to the infinite family of eigenmodes for the Euler buckling of a slender rod. Special attention is paid to the catenoid, which emerges as the shape of maximal allowable separation when the area is less than a critical area equal to the planar area enclosed by the two rings. A perturbation theory argument directly relates the tension of catenoidal membranes to the stability of catenoidal soap films in this regime. When the membrane area is larger than the critical area, we find additional cylindrical tether solutions to the shape equations at large ring separation, and that arbitrarily large ring separations are possible. These results apply for the case of vanishing Gaussian curvature modulus; when the Gaussian curvature modulus is nonzero and the area is below the critical area, the force and the membrane tension diverge as the ring separation approaches its maximum value. We also examine the stability of our shapes and analytically show that catenoidal membranes have markedly different stability properties than their soap film counterparts.
} \\

\end{tabular}

 \end{@twocolumnfalse} \vspace{0.6cm}

  ]

\renewcommand*\rmdefault{bch}\normalfont\upshape
\rmfamily
\section*{}
\vspace{-1cm}


\footnotetext{\textit{$^{a}$~Flatiron Institute, Center for Computational Biology, 162 5th Avenue, New York, NY 10010, USA. E-mail: ljia@flatironinstitute.org}}
\footnotetext{\textit{$^{b}$~Theoretical Physics Center and Department of Physics, Brown University, Providence, RI 02912 USA. }}
\footnotetext{\textit{$^{c}$~Center for Fluid Mechanics and School of Engineering, Brown University, Providence, RI 02912, USA. }}

\footnotetext{\dag~Electronic Supplementary Information (ESI) available. See DOI: 00.0000/00000000.}

\section{Introduction}

\begin{figure}[t!]
\centering
\includegraphics[height=2.5in]{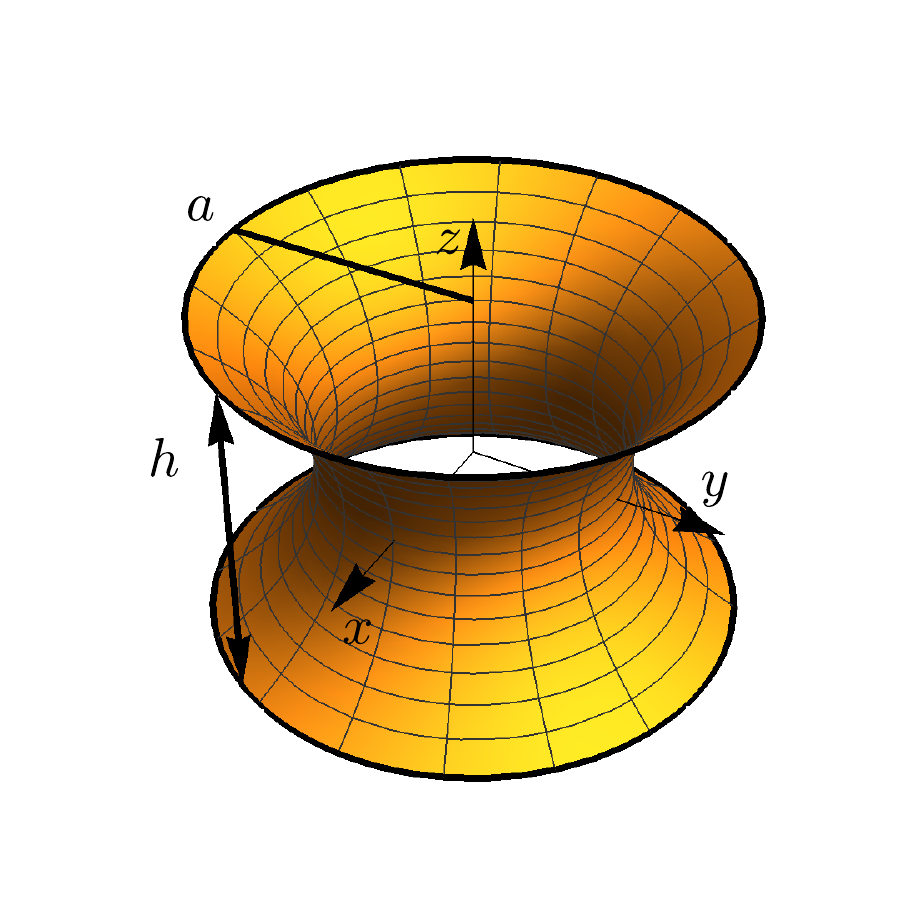}
\caption{(Color online.) A membrane connecting two rings of radius $a$, separated by distance $h$. The surface is a surface of revolution described by $r(z)$. The shape of this particular membrane is the catenoid of greatest area for the ring radius $a$; or equivalently, the catenoid with greatest $h$ for the given $a$.}
\label{fig:critcat}
\end{figure}

Although the lowest energy state of a symmetric biological membrane is flat, membranes in the cell can be curved because of forces external to the membrane, such as the forces arising from scaffolding proteins or the cytoskeleton~\cite{JarschDasteGallop2016}. 
In this paper we consider a simple idealized problem for determining how membrane shape depends on external force. We study a fluid membrane of fixed area connected to two rings (Fig.~\ref{fig:critcat}). The area is fixed because bending a thin membrane is much easier than stretching it. The rings are parallel, have the same radius, and have aligned centers. This setup is similar to that used to study the catenoid formed by a soap film~\cite{CryerSteen1992,RobinsonSteen2001,Salkin_etal2015,Goldstein_etal2021} or a smectic film stretched between two rings~\cite{BenAmar_etal1998,Chikina_etal1998,MullerStannarius2006,MayHarthTrittelStannarius2012}, or a capillary bridge~\cite{GilletteDyson1971,OrrScrivenRivas1975}.  A similar setup has also been used to study membrane tethers at fixed tension~\cite{HeinrichWaugh1996,powers_huber_goldstein2002,DerenyiJulicherProst2002}.
The membrane has two circular edges connected to the rings. The rings exert zero torque on the membrane edge. But since the rings have a fixed radius and exert the force required to obtain a given membrane extension, the membrane edges are not completely free as in the case of lipid bilayer membranes with reduced edge tension~\cite{fromherz1983lipid,SaitohTakiguchiTanakaHotani1998,zhao2005monte}, or colloidal membranes comprised of rod-like viruses~\cite{BarryBellerDogic2009,BarryDogic2010,Gibaud_etal2012}.
In general, the external force has a dominant effect on the shape. In the absence of the external force,  many of the simplest possible  surfaces with edges are ruled out for a membrane with bending stiffness~\cite{Tu2010,Tu2011}. The condition of zero force and zero torque at the edge rules out surfaces with edges and constant mean curvature like a cylinder, a catenoid, an unduloid~\cite{Delaunay1841}, or part of a sphere; a surface which is part of the Willmore torus~\cite{Willmore1965,Willmore1993}; or a surface which is part of biconcave discoid shape. We will see that some of these shapes are allowed when there is an external force. The scope of this paper is limited to axisymmetric shapes, although some non-axisymmetric shapes such as helicoids can be treated by similar methods~\cite{Balchunas_etal2019b}. 

Our work is complementary to recent work in the mathematics community on the  shapes that are critical points of the bending energy, such as the study of axisymmetric shapes with zero mean curvature at the edges and with no constraint on the area~\cite{DallAcquaDeckelnickWheeler2013}, or the study of axisymmetric shapes with fixed tension~\cite{DeckelnickDoemelandGrunau2021}. The paper of Deckelenick and Grunau~\cite{DeckelnickGrunau2009} is an important precursor for our present article since their numerical experiments suggest a rich collection of possible shapes in the case of no area constraint and vanishing mean curvature at the edges of the surface. Our work is distinct from these investigations since we enforce the constraint of fixed area  and impose the most general condition of vanishing bending moment at the edge, i.e. with nonzero Gaussian curvature modulus.

We begin our analysis in Sec.~\ref{soap} with a review of the properties of the catenoid in the context of the soap film problem. Then in Sec.~\ref{Willmore} we review the Willmore problem, which is to find the shape that minimizes the integral of the square of the mean curvature without a constraint on the area. Part of the Willmore torus turns out to be one of the solutions to our problem at a certain area and ring separation. Section~\ref{Sec:membrane_Eqs} sets the notation we use for the standard Canham-Helfrich energy for a membrane with fixed area, as well the parametrization for axisymmetric shapes. In Sec.~\ref{Sec:Results} we present our main results, showing that there are three regimes of behavior depending on the area.  We begin in Sec.~\ref{zeroGK} with the case of zero Gaussian curvature modulus. For small area, we find two solutions for each extension below a maximum extension, at which a catenoid forms. There is a regime of intermediate area for which two catenoids are allowed at two specific values of the extension, as well as extended `tether' shapes which may be drawn out to arbitrary length. At the greatest areas, no catenoids ever form, but tethers form at large extension.
In the rest of Sec.~\ref{Sec:Results} we consider the case of nonzero Gaussian curvature modulus; some special isolated shapes such as spheres, cylinders, and Willmore tori; and  stability.
Section~\ref{discuss_and_conclude} is the conclusion. An appendix summarizes the differential geometry formulas we use and reviews the argument that the Noether invariant for this problem is the axial force. 

\subsection{Soap film problem: zero mean curvature}\label{soap} 

\begin{figure}[h]
\centering
\includegraphics[height=2.6in]{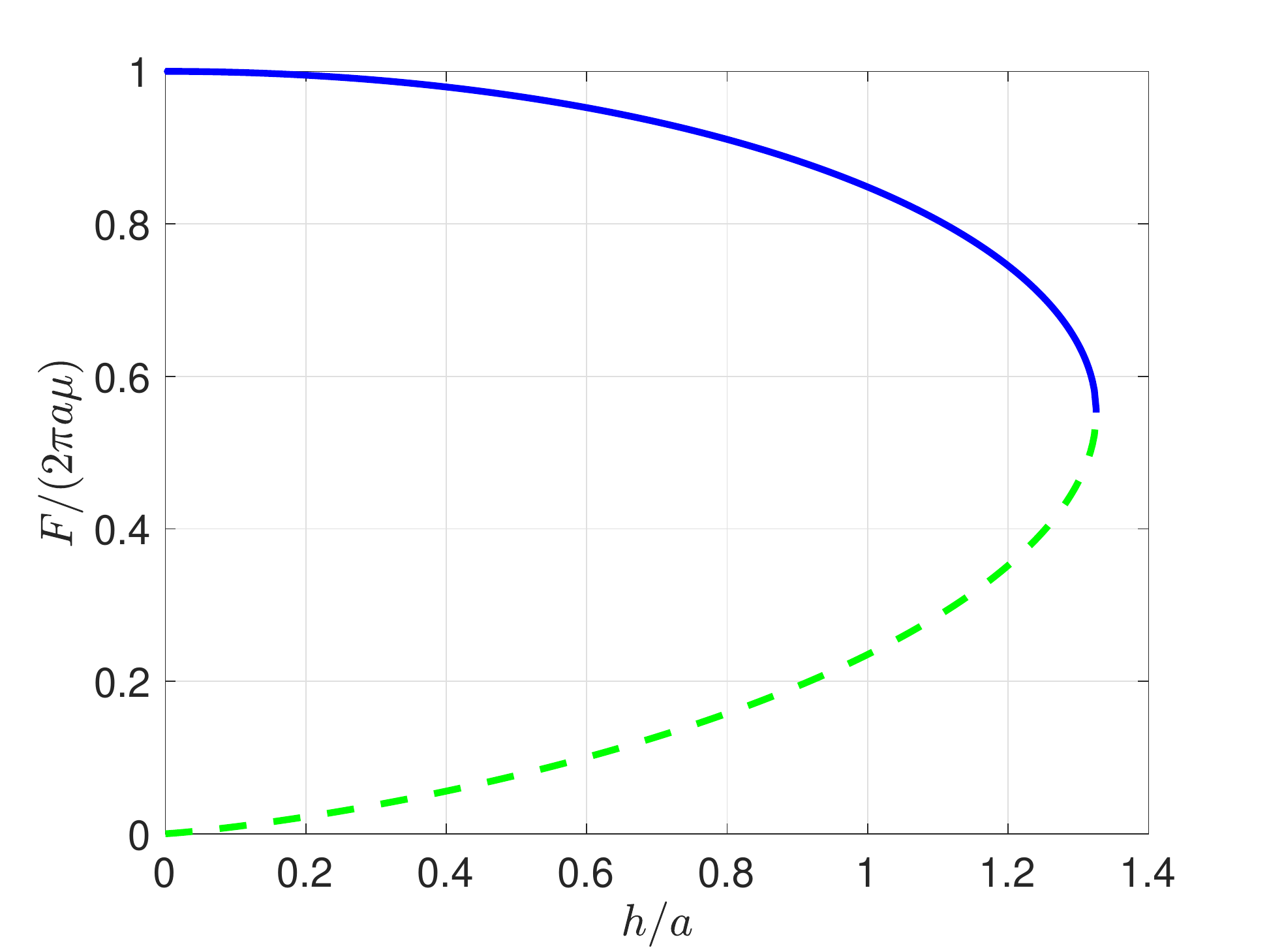}
\caption{
Dimensionless force $F/(2\pi a\mu)$ vs. dimensionless height $h/a$ of the catenoid. The upper solid branch (blue) corresponds to shapes with lower area, for a given separation $h$, than the lower dashed branch (green).}
\label{fig:Fvh}
\end{figure}

First we review the classic problem of a soap film stretched between two rings. The rings each have radius $a$ and are separated by a distance $h$. The rings are parallel to each other and lie in planes normal to the $z$ axis, with centers on the $z$ axis (Fig.~\ref{fig:critcat}). Since the energy of the soap film is the surface tension $\mu$ times the area, the equilibrium shape minimizes the area. The condition for the surface to be an extremum of area, or more simply a minimal surface, is that the mean curvature vanishes~\cite{struik1988}. In our convention a sphere has negative mean curvature; the basic formulas are summarized in Appendix~\ref{dgappx}. In cylindrical coordinates in which radius $r$ is a function of $z$ and we denote derivatives with respect to $z$ via subscripts, this condition is 
\begin{equation}
\frac{1}{r \sqrt{1+r_z^2}}-\frac{r_{zz}}{(1+r_z^2)^{3/2}}=0.
\end{equation}
The catenoid,
\begin{equation}
r=b\cosh(z/b),\label{eqncat}
\end{equation} is the only nonplanar surface of revolution that is a minimal surface~\cite{FomenkoTuzhilin1991}.  The parameter $b$ is the radius of the neck of the catenoid; it is related to the ring separation $h$ by $h=2b\cosh^{-1}(a/b)$. The force required to hold the rings apart at fixed separation is given by $F=2\pi\mu b$. Figure~\ref{fig:Fvh} shows the force as a function of ring separation~\cite{BenAmar_etal1998}, and reveals that as long as the separation is less than the maximum value of separation $h_\mathrm{max}\approx1.3255a$, there are two catenoids connecting the rings. There is one catenoid solution at  $h=h_\mathrm{max}$, and no catenoid solutions for $h>h _\mathrm{max}$. 

\begin{figure}[h!]
\centering
\includegraphics[height=2.6in]{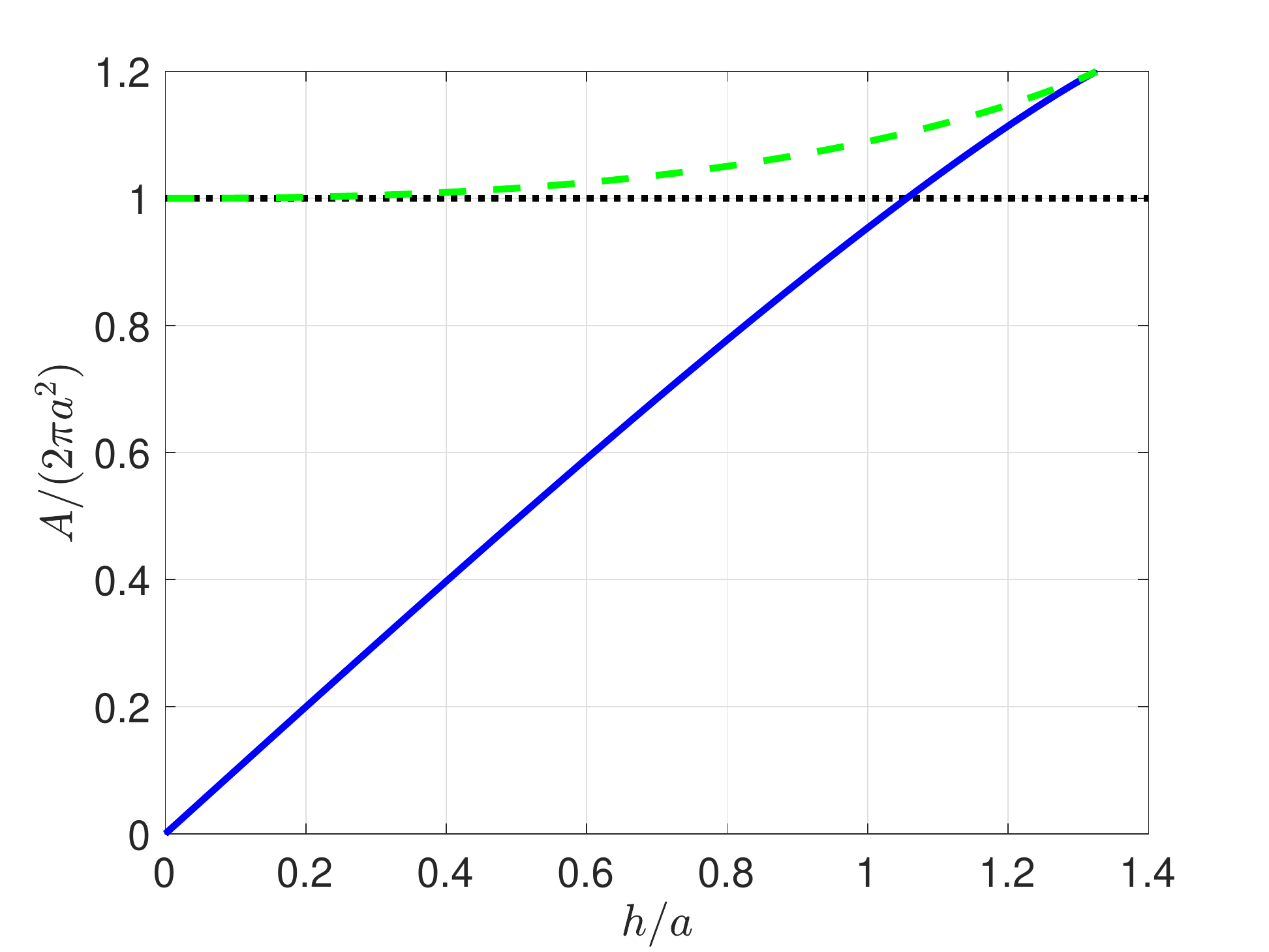}
\caption{
Dimensionless area $A/(2\pi a^2)$ vs. dimensionless height $h/a$ of the catenoid. The color scheme is the same as in Fig.~\ref{fig:Fvh}: for a given separation $h/a$, the catenoid with larger neck radius $b$ has smaller area. The dotted line corresponds to the area of the Goldschmidt solution, in which a flat membrane spans each ring. The dimensionless area $\bar{A}$ of the critical catenoid at the maximal extension is $\bar{A} = \bar{A}_\mathrm{max}\approx1.1997$.
}
\label{fig:Avh}
\end{figure}

The area of a catenoid of neck radius $b$ and ring separation $h$ is given by $A=\pi b[h+b\sinh(h/b)]$. For a given $h/a$, the catenoid with the larger neck radius has less area (Figs.~\ref{fig:Fvh} and~\ref{fig:Avh}).
This branch of catenoids is also stable to small perturbations, whereas the larger-area branch is unstable~\cite{Chikina_etal1998,BenAmar_etal1998}. See ref.~\cite{Chikina_etal1998} for a photograph of the stable and unstable catenoids for the same $h$.
Note that for a given ring radius $a$, the critical catenoid at the largest extension $h\approx1.3255a$ is also the catenoid of greatest area~\cite{TaylorMichael1972} 
(Fig.~\ref{fig:Avh}), with $\bar{A}\equiv A/(2\pi a^2)\approx 1.1997$, a number we define to be $\bar{A}_\mathrm{max}$. {We henceforth refer to the catenoid with larger neck as the ``thick catenoid'' and the catenoid with smaller neck as the ``thin catenoid.''} 

While the catenoid locally minimizes area among continuous surfaces of revolution, it is not necessarily an absolute minimum. Consider the discontinuous Goldschmidt solution consisting of two disks of radius $a$ with center-to-center distance $h$. Regularizing this solution by adding a thin connecting cylinder produces a continuous shape of area $A \to 2\pi a^2$ as the radius of the connecting cylinder vanishes. In particular, when $\bar{A} > 1$, 
the Goldschmidt solution has less area than both the thin and thick catenoids.

\subsection{Willmore problem}\label{Willmore}
\begin{figure*}[t]
\centering
\includegraphics[height=2.2in]{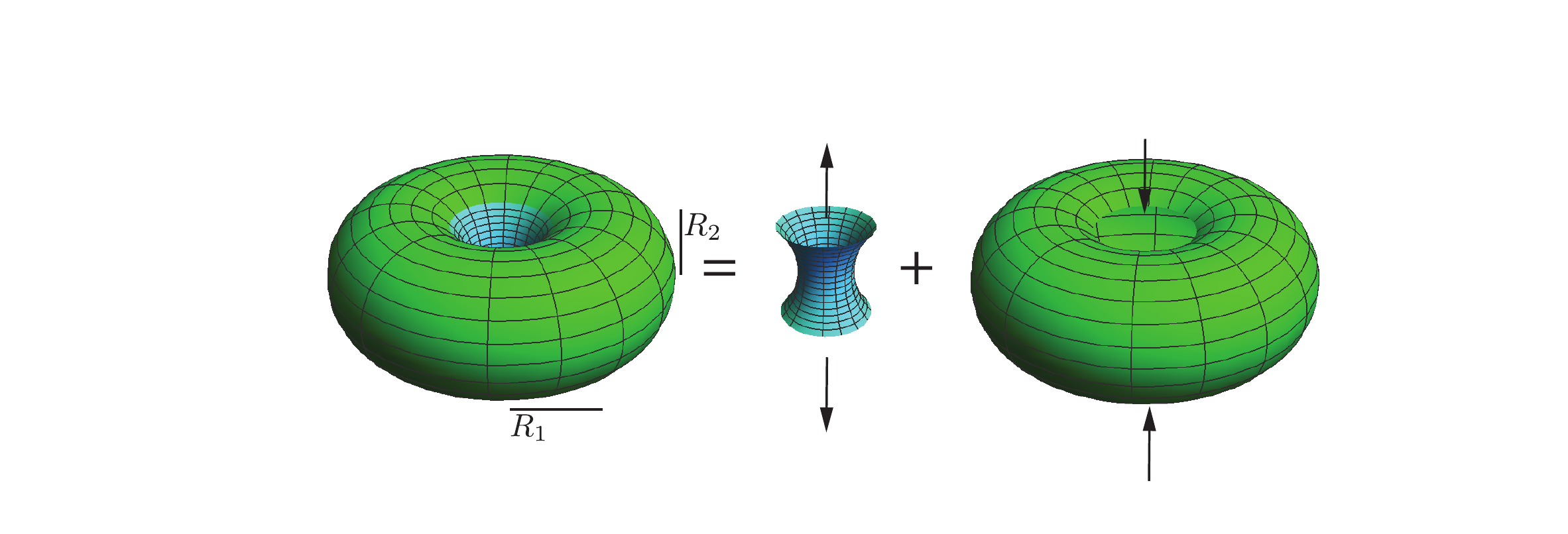}
\caption{ A Willmore torus, which has $R_2/R_1=1/\sqrt{2}$, (left) can be cut to make two axisymmetric shapes of different area. The torus is force-free, but since there are internal bending forces in the torus, the cut shapes require external forces along their boundary circles to maintain equilibrium. The net forces are indicated by the arrows; the  force on the inner part of the torus (middle) is tensile, equal in magnitude to the compressive force on the outer part of the torus (right). The circular cuts in this figure are made along the curves with vanishing mean curvature; the green color indicates that the mean curvature is negative, whereas the blue color signifies positive mean curvature. }
\label{fig:torus}
\end{figure*}
Next we consider a surface with a cost for bending only, but with no constraint on the area.  A classic mathematical problem is to find the surface of given topological character that has the least possible curvature, as measured by the bending energy
\begin{equation}
E_\mathrm{W}=\int\mathrm{d}AH^2,\label{WillE}
\end{equation}
where $H$ is the mean curvature and $\mathrm{d}A$ is the area element (see~\cite{PinkallSterling1987} and~\cite{Nitsche1993} for surveys). Willmore showed that the energy $E_\mathrm{W}$ satisfies $E_\mathrm{W}\ge4\pi$ for any closed orientable surface~\cite{Willmore1965,Willmore1993}. It is easily checked that any sphere gives the minimum energy.
Note that the bending energy eqn~(\ref{WillE}) is invariant under conformal transformations of three-dimensional space~\cite{White1973,Blaschke1929}. Willmore further considered the case of tori, and showed that $E_\mathrm{W}\ge2\pi^2$ for the special class of tori formed by a tube of constant radius around a closed space curve. He showed that the torus of this type that gives the minimum energy is the one in which the space curve is a circle and the radius of the tube is $1/\sqrt{2}$ times the radius of the circle (Fig.~\ref{fig:torus}, left), and conjectured that this torus minimizes $E_\mathrm{W}$ over all surfaces with the topology of the torus~\cite{Willmore1965,Willmore1993}. The Willmore conjecture was shown to be true by Marques and Neves~\cite{MarquesNeves2014}.

To connect this problem with the problem of stretching a membrane between two rings, suppose we cut the Willmore torus along circles of unit radius to make two surfaces (Fig.~\ref{fig:torus}).

The Euler-Lagrange equation for the energy $E_\mathrm{W}$ is
\begin{equation}
\Delta H+2H(H^2-K)=0,\label{WillmoreEq}
\end{equation}
where $\Delta$ is the Laplacian and $K$ is the Gaussian curvature~\cite{zhong-can_helfrich1989}.
Any minimal surface, such as a catenoid, satisfies the Euler-Lagrange equation since it has $H=0$. Spheres also 
satisfy eqn~(\ref{WillmoreEq}) since the mean curvature is uniform and $H^2=K$.
Now consider a torus formed by a tube of radius $R_2$ with the centerline of the tube a circle of radius $R_1$:
\begin{equation}
r(z)=R_1\pm\sqrt{R_2^2-z^2}
\end{equation} 
This surface is not a minimal surface,  but it satisfies eqn~(\ref{WillmoreEq}) when it is a section of a Willmore torus, i.e. $R_2=R_1/\sqrt{2}$.
We will review below how to calculate the force required to hold in equilibrium a surface with the bending energy eqn~(\ref{WillE}), but the inner part of the torus is under tension, while the outer part is under compression. We'll also see that the bending moment acting at the edge is given by the mean curvature $H$; in Fig.~\ref{fig:torus} we chose to cut the torus along the two circles that have $H=0$ so that no bending moment is required in equilibrium. 
It was shown by Deckelnick and Grunau~\cite{DeckelnickGrunau2009} that this solution is not an isolated solution but part of a family of solutions. 
Our  constraint of fixed area leads to a different set of solutions.


\section{Membrane equations}
\label{Sec:membrane_Eqs}

Next, we turn to the problem of a membrane that resists bending at fixed area. The approximation of fixed area is valid as long as the tension is small compared to the area expansion modulus. Our goal is to calculate the shape of an axisymmetric membrane of fixed area connecting two circular rings.  We also calculate the force as a function of ring displacement. 

\subsection{Governing equations}
We assume the energy of the membrane is given by the Canham-Helfrich energy with a {Lagrange multiplier $\mu$ corresponding to the tension and enforcing the constraint of fixed area}:
\begin{equation}
E=\frac{\kappa}{2}\int\mathrm{d}A (2H)^2+\bar{\kappa}\int\mathrm{d}A K+\mu\int\mathrm{d}A.
\label{EEq}
\end{equation}
This energy is a simple generalization of the Willmore energy $E_W$ of eqn~(\ref{WillE}),
with $\kappa$ the bending modulus,  and $\bar{\kappa}$ the Gaussian curvature modulus~\cite{canham1970,helfrich1973}. 
Motivated by recent work on colloidal membranes~\cite{Gibaud_etal2017}, we study the case of a positive Gaussian curvature modulus.  Mathematically a positive Gaussian curvature can be problematic since it favors arbitrarily large negative $K$; therefore, some authors~\cite{Scholtes2011} consider the case of a negative Gaussian curvature modulus. Sometimes higher order terms must be introduced to stabilize the system when the Gaussian curvature modulus is positive~\cite{KaplanTuPelcovitsMeyer2010}. In our problem, the penalty for mean curvature and the area constraint prevent the the Gaussian curvature from becoming arbitrarily large.

The Euler-Lagrange equation is given by~\cite{zhong-can_helfrich1989}
\begin{equation}
\kappa\left(\Delta H+2H^3-2HK\right)-\mu H=0,\label{shapeEq}
\end{equation}
which differs from eqn~(\ref{WillmoreEq}) only by the term linear in the mean curvature arising from the area constraint. 
The condition of vanishing bending torque at the edge  is~\cite{CapovillaGuvenSantiago2002,TuOu-Yang2004} 
\begin{equation}
2\kappa H+\bar{\kappa} k_\mathrm{n}=0,\label{notorque}
\end{equation}
where $k_\mathrm{n}$ is the normal curvature of the boundary. 
If $\hat{\mathbf{T}}$ is the tangent vector of the boundary 
and $\hat{\mathbf{n}}_C$ is the  membrane normal on the boundary, then $k_\mathrm{n}=\hat{\mathbf{n}}_C\cdot\mathrm{d}\hat{\mathbf{T}}/\mathrm{d}l$, where $l$ is arclength along the boundary. The convention is that $l$ is increasing when the surface is on the left of the boundary.
Note that for our circular boundaries, $k_\mathrm{n}<0$.


\begin{figure}[h!]
\centering
\includegraphics[height=6cm]{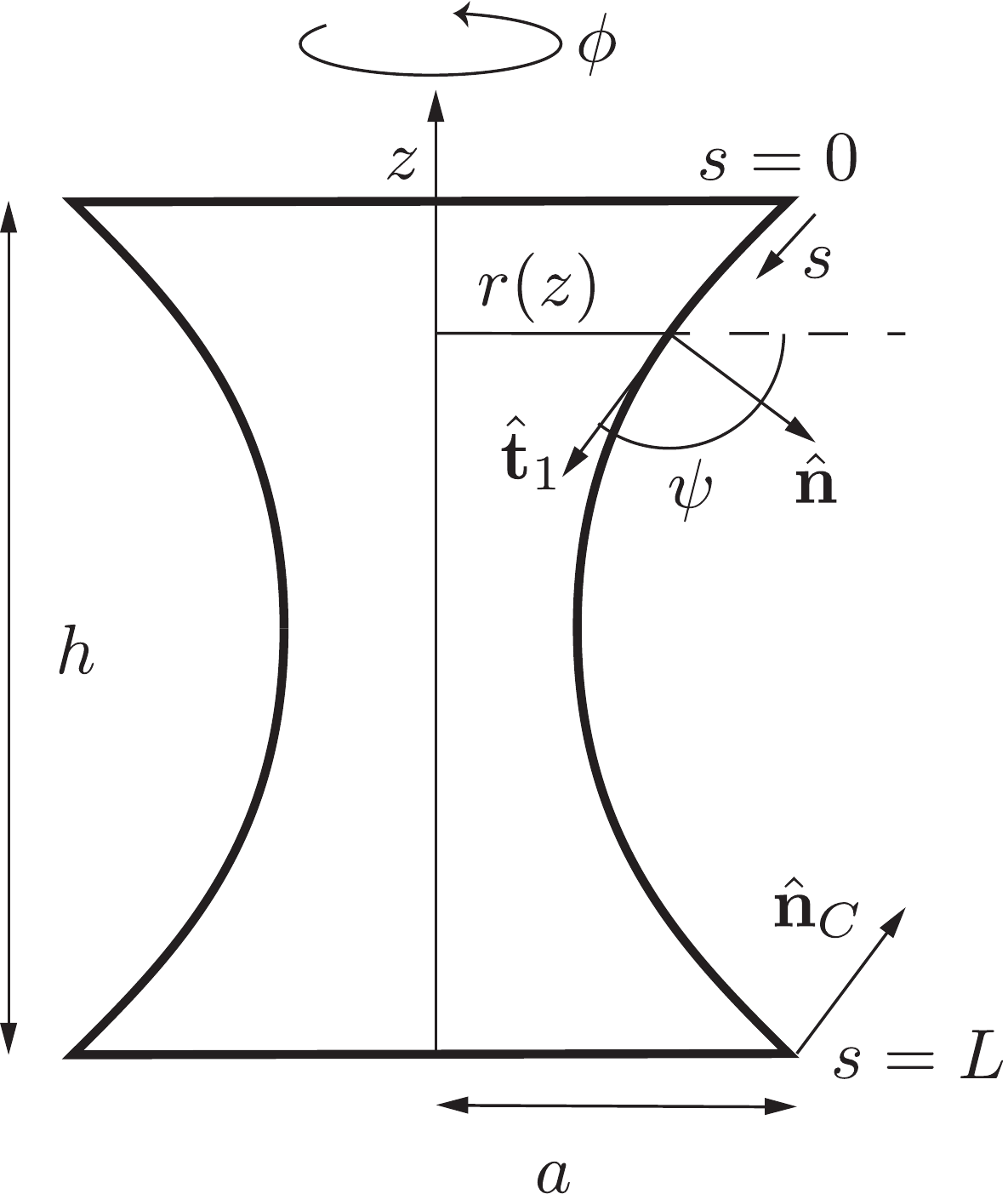}
\caption{The surface of revolution obtained by rotating the graph $r(z)$ about the $z$-axis. The surface connects two parallel circular rings of radius $a$, separated by distance $h$. The surface is parameterized by arclength $s$ along a meridian. The normal to the surface is $\hat{\mathbf n}$ and the normal  to the surface at the boundary is $\hat{\mathbf n}_C$. The angle $\psi$ is measured from the outward radial direction to the surface tangent
, and the length of the meridian is $L$.}
\label{fig:setup}
\end{figure}

\subsection{Parameterization}\label{parameterization}
We follow the approach of J\"ulicher and Seifert~\cite{JulicherSeifert1994}, denoting the contour of the membrane by $r$ and $z$, which are functions of the arclength $s$ measured along the contour (Fig.~\ref{fig:setup}). The angle $\psi$ is the angle between the contour tangent vector $\mathbf{t}_1=r_s\hat{\mathbf r}+z_s\hat{\mathbf z}$ and the radial direction $\hat{\mathbf r}$, so that 
\begin{eqnarray}
r_s&=&\cos\psi\\
z_s&=&-\sin\psi.
\end{eqnarray}
Our task is to minimize the energy {$E$, eqn~(\ref{EEq}). Writing $E=\int\mathrm{d}s\mathcal{E}$, we introduce the energy density  $\mathcal{E}$:} 
\begin{eqnarray}
\frac{\mathcal{E}}{2\pi}&=&\frac{\kappa}{2}(2H)^2r+\bar{\kappa} r K
+\mu r\nonumber\\
&+&\left[\gamma(s)(r_s-\cos\psi)
+\eta(s)(z_s+\sin\psi)\right],
\end{eqnarray}
where
$2H=-[\psi_s+(\sin\psi)/r]$, $K=(\psi_s\sin\psi)/r$, and we have introduced the $s$-dependent  Lagrange multipliers $\gamma$ and $\eta$ to allow the variations in $r$, $\psi$, and $z$ to be taken independently (see the appendix for definitions of the geometrical quantities). Note that the boundary term arising from the variation of $z$ is $2\pi\eta\delta z|^L_0$; in other words, the axial force required to hold the rings with separation $h$ is 
\begin{equation}
\label{forceetaeqn}
F=-2\pi\eta.
\end{equation}
We follow the standard procedures of variational calculus when the end points of the domain are free to move~\cite{GelfandFomin1963}, since although value of the arclength $s$ is fixed at $s=0$ at one endpoint, the value of $s$ at the other endpoint, $s=L$, is only determined once the problem is completely solved.  The variation of the end point at $s=L$ leads to the boundary condition~\cite{JulicherSeifert1994} 
$\mathcal{H}(s=L)=0$, where $\mathcal{H}$ is the Hamiltonian obtained from the Legendre transform of $\mathcal{E}$:
\begin{equation}
\mathcal{H}=\psi_s\frac{\partial\mathcal{E}}{\partial\psi_s}+r_s\frac{\partial\mathcal{E}}{\partial r_s}+z_s\frac{\partial\mathcal{E}}{\partial z_s}-\mathcal{E}.
\end{equation}
In our parameterization,
\begin{eqnarray}
\frac{\mathcal{H}}{2\pi}&=& \frac{\kappa}{2}\left[r\psi_s^2-\frac{\sin^2\psi}{r}\right]-\mu r+\gamma\cos\psi-\eta\sin\psi\\
&=&-\kappa rH\left[\psi_s-\frac{\sin\psi}{r}\right]-\mu r+\gamma\cos\psi-\eta\sin\psi.
\end{eqnarray}
Note that $\mathcal{H}=0$ for all $s$, since it is the conserved quantity associated with the fact that the arclength $s$ does not appear explicitly in $\mathcal{E}$. The other boundary conditions are $r(0)=r(L)=a$, $z(0)=h$, $z(L)=0$, and the condition of vanishing bending moment at either ring:
\begin{equation}
\left[\kappa\left(r \psi_s+\sin\psi\right)+\bar{\kappa}\sin\psi\right]_{s=0,s=L}=0.
\end{equation}
Defining $t=\tp{-1/2}+s/L$ and denoting derivatives with respect to $t$ with a dot, the Euler-Lagrange equations are~\cite{JulicherSeifert1994}
\begin{eqnarray}
\dot{\psi}&=&L\psi_s\label{psi-dot-eq}\\
\frac{r}{L}\dot{\psi_s}&=&- \cos\psi\psi_s+\frac{\sin\psi\cos\psi}{r}\nonumber\\
&&+\frac{\gamma}{\kappa}\sin\psi+\frac{\eta}{\kappa}\cos\psi\\
\dot{r}&=&L\cos\psi\\
\dot{z}&=&-L\sin\psi\\
\dot{\gamma}&=&L\left[\frac{\kappa}{2}\psi_s^2-\frac{\kappa}{2}\frac{\sin^2\psi}{r^2}+\mu\right]\\
&=&L\left[-\kappa H\left(\psi_s+\frac{\sin\psi}{r}\right)+\mu\right]\\
\dot{\eta}&=&0.\label{eta-dot-eq}
\end{eqnarray}
To these equations we add the area equation $\dot{\mathcal{A}}=2\pi rL$, where $\mathcal{A}\equiv2\pi\int_{\tp{-1/2}}^t\mathrm{d}t^\prime L r(t)$. We also add
the conditions that $\mu$ and $L$ are constant, leading to a total of nine first-order equations. In addition to the seven boundary conditions we have already mentioned [for the quantities $r(t=\tp{-1/2})$, $r(t=\tp{1/2})$, $z(t=\tp{-1/2})$, $z(t=\tp{1/2})$, the bending moment at either endpoint, and $\mathcal{H}(t=1)$], we add the conditions on the area function: $\mathcal{A}(t=\tp{-1/2})=0$, and $\mathcal{A}(t=\tp{1/2})=A$, where $A$ is the imposed area.



\section{Results}\label{Sec:Results}


Solving the Euler-Lagrange equations in the geometry described above reveals a whole zoo of axisymmetric shapes, both familiar and unfamiliar. We blend analytical and numerical approaches to probe the shapes that form in various parameter regimes.
Given an extension $h$ and an area $A$, the MATLAB routine \texttt{bvp4c} was used to numerically solve eqn~(\ref{psi-dot-eq})-(\ref{eta-dot-eq}) and their associated boundary conditions. By treating $\mu$ as an additional dependent variable in these equations, the tension can be computed as part of this procedure. The axial force is calculated from eqn~ (\ref{forceetaeqn}). We give an equivalent, perhaps more physical expression for the axial force in terms of the tension and bending stiffness in eqn~(\ref{forceEq}) of Appendix A.2. Once a solution at extension $h$ is obtained, it is used as an initial guess for the solution at a nearby extension $h + \delta h$; in this way, we determined shapes, tension, and force as functions of extension. Strictly speaking, negative extensions are  numerically permissible, but since negative extension requires the two boundary rings to pass through each other, we generally do not concern ourselves with this unphysical regime. The same goes for self-intersecting solutions, which can occur when the extension becomes small.

\begin{figure*}[t]
\centering
\includegraphics[scale=0.115]{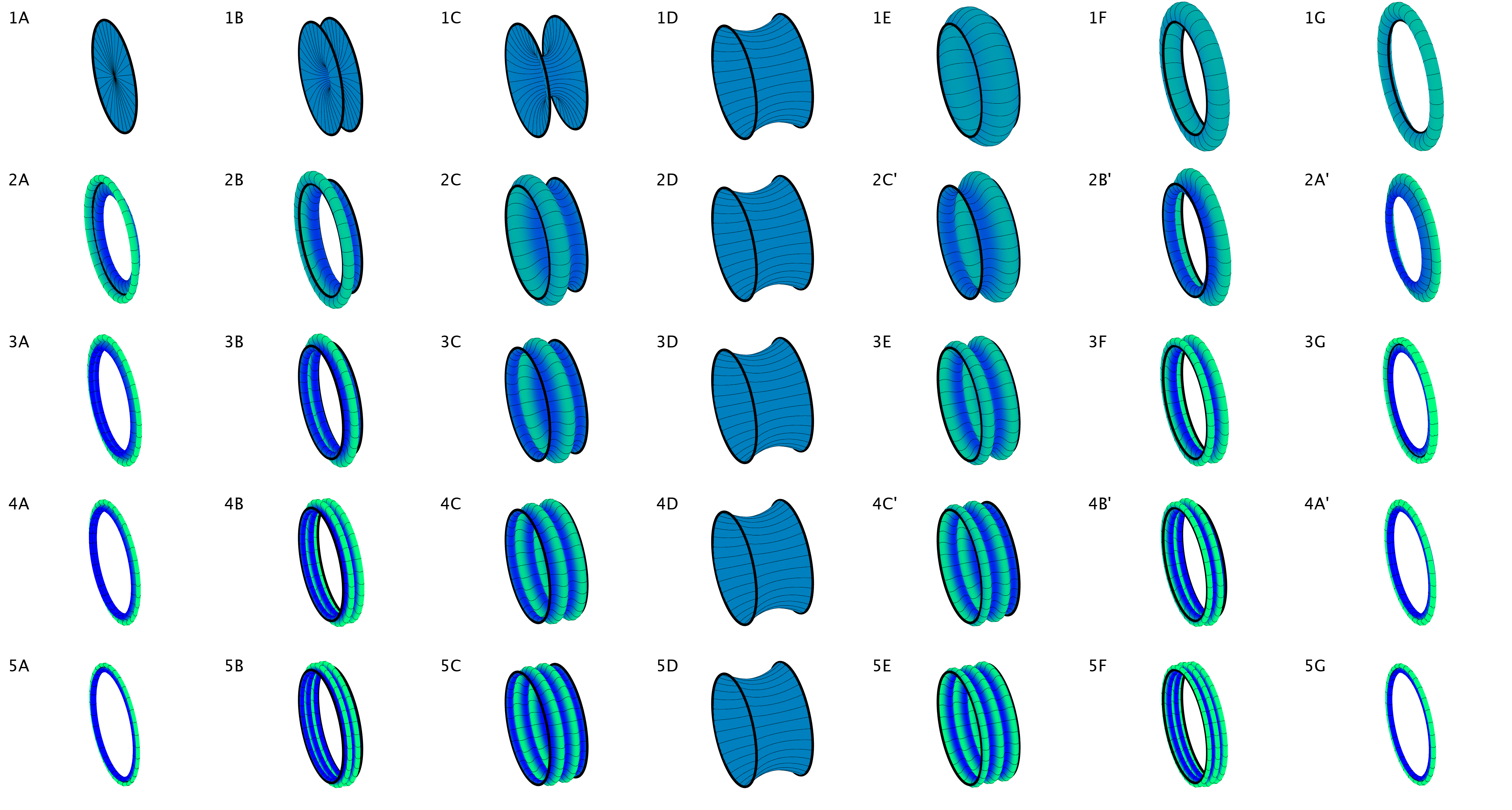}
\includegraphics[scale=0.25]{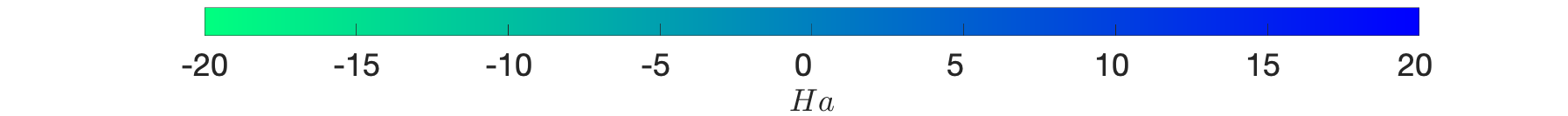}
\caption{The first five buckling modes of an axisymmetric membrane with $\bar\kappa = 0$ and dimensionless area $\bar{A} = 1$. Each row of shapes corresponds to a single mode, first traversing the branch of less negative force with extensions $h/a = 0$ (A), $h/a = 0.35$ (B), $h/a = 0.7$ (C), $h/a \approx 1.0554$ (D, the catenoid), then continuously traversing the branch of more negative force with extensions $h/a = 0.70$ (E), $h/a = 0.35$ (F), and $h/a = 0$ (G). 
Branch behavior depends on the parity of $n$: if $n$ is odd, the shapes on each branch are distinct, and if $n$ is even, the shapes on each branch are mirror images of each other (denoted by a prime). Surfaces can self-intersect as $h/a$ approaches zero. All shapes shown here require a compressive stabilizing force at the boundary rings. Color indicates the dimensionless local mean curvature $Ha$. Black rings of radius $a$  have been added at $z=\pm h/2$ for visibility. }
\label{fig:catshapes}
\end{figure*}



\subsection{Case of zero Gaussian curvature modulus} \label{zeroGK}
We begin by considering the simplified case of $\bar\kappa = 0$. In this case, the condition of vanishing torque at the edge, eqn~(\ref{notorque}), implies that the mean curvature vanishes at the edge. 
Our first observation is that there are three distinct parameter regimes governed by area. To begin, suppose the area of the membrane is less than the area of the planar disks bounded by the two rings, $\bar{A} \leq 1$. In this regime, the axisymmetric surface of least area is always the thick catenoid. A quick argument shows that the membrane has a finite maximal extension $h^* =h^*(A)$, where $h^*(A)$ is the separation of the rings for a thick catenoid of area $A$ (the solid blue curve in Fig.~\ref{fig:Avh}). A hypothetical shape with $h>h^*$ would have area strictly less than that of the thick catenoid with separation $h$, which is by definition area minimizing. Furthermore, the shape of the membrane when $h=h^*$ will always be a thick catenoid because this shape  is the unique axisymmetric surface that can be formed for the given extension and area.

The second regime is when $1 < \bar{A} < \bar{A}_\mathrm{max}$, where $\bar{A}_\mathrm{max}\approx1.1997$ is the critical dimensionless area beyond which catenoids do not exist. 
According to Fig.~\ref{fig:Avh}, this is the regime where there are two possible catenoids for the given area, one thick and one thin. Essentially, the number of branches is doubled, and each branch has a maximal extension that again corresponds to the width of one of the two catenoids. The notable exception in this regime is the tether which has no maximal extension; mathematically, this is a regularized Goldschmidt solution, which is possible when the area exceeds that of two disks.

The final regime, $\bar{A} \geq \bar{A}_\mathrm{max}$, is where the area is sufficiently large that catenoids do not appear at all. Here, the shapes of maximal extension are previously unknown non-catenoidal surfaces. Unlike the previous cases, the branches of the thick and thin (former) catenoids meet each other at these shapes of maximal extension. The tether remains a possible solution in this regime as well.

\subsubsection{Small area: $\bar{A}\leq   1$}
\label{smallareasec}

As argued above, when the dimensionless area $\bar{A}$ is smaller than the critical value of 1, we need only consider extensions in the range $0\leq h\leq h^*$, with the catenoid known to be the equilibrium shape at maximal extension $h=h^*$. This characterization of the catenoid leads to a curious phenomenon: suppose we wish to calculate the tension of the catenoid. Directly substituting $H=0$ into the Euler-Lagrange equations does not lead to a form where the tension can be calculated by applying the area constraint; instead, $\mu$ drops out and is left undetermined.

\begin{figure}[t!]
\centering
\includegraphics[scale=0.2]{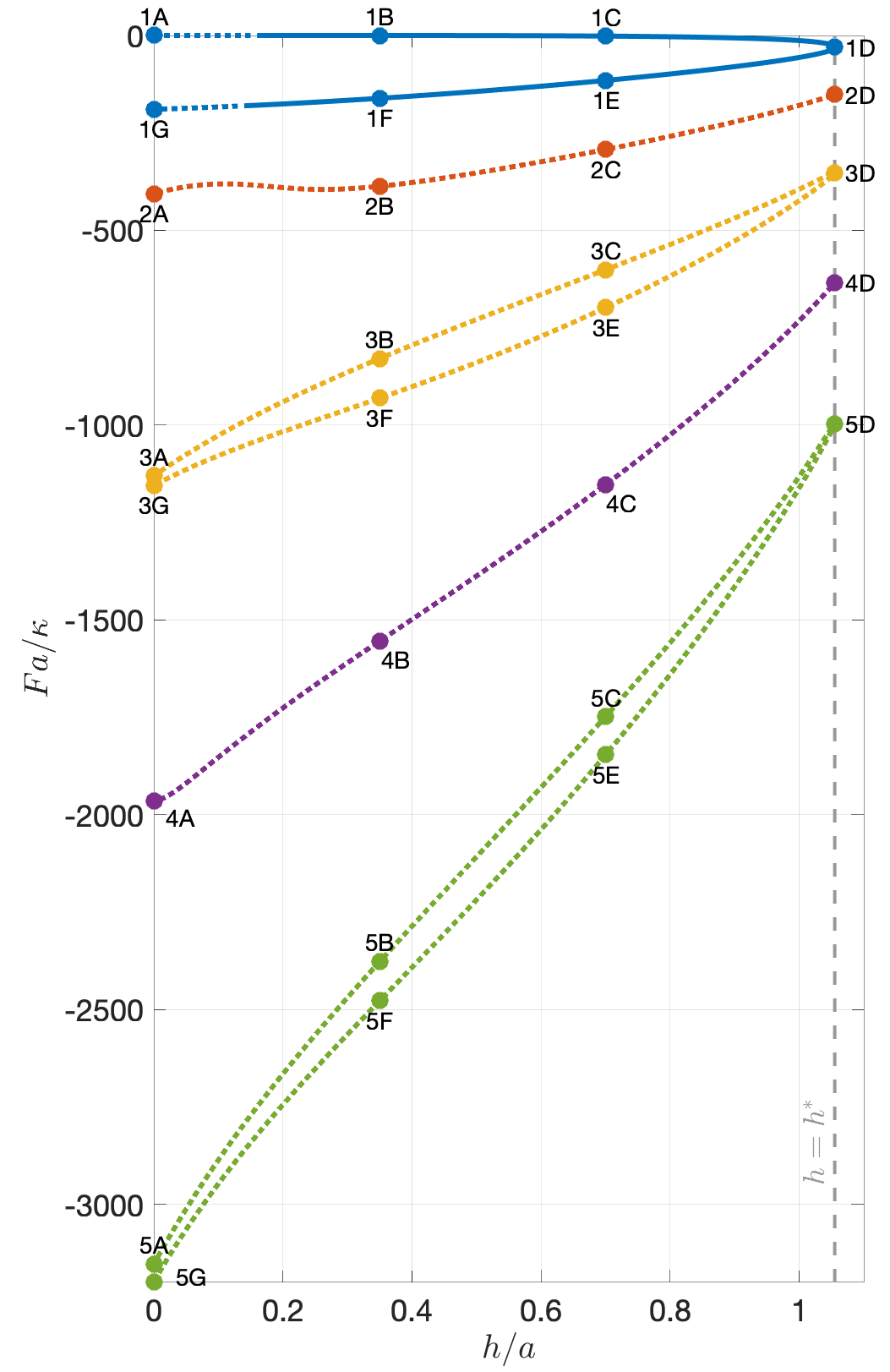}
\caption{Dimensionless force vs. extension for the first five bending modes of a membrane with $\bar{A} =1$ and $\bar\kappa = 0$.  Labeled points correspond to the shapes in Fig.~\ref{fig:catshapes}. Each branch has a maximal extension $h^*/a\approx1.0554$ (grey dashed line), where the membrane assumes the shape of a catenoid with compressive force $F = 2\pi \mu b$ and beyond which no equilibrium surfaces exist. For even modes, the shapes of the two branches are mirror images of each other and hence their curves in the force vs. extension diagram overlap; for odd modes, the two curves are similar but different. Stable surfaces are indicated with a solid line while unstable surfaces are indicated with a dotted line; for $\bar{A}=1$ and $\bar\kappa =0$, the only stable surfaces are found at the lowest mode. }
\label{fig:Fvh_smallA}
\end{figure}

To circumvent this issue, we instead formulate the calculation by perturbing the Euler-Lagrange equation, eqn~(\ref{shapeEq}), around the catenoid state. From the formula for the catenoid, eqn~(\ref{eqncat}), we find that the arclength is given by $L_0/2-s=b\sinh(z/b)$, where $L_0$ is the contour length of a longitude of the catenoid. 
Evaluating $\cosh(z/b)^2-\sinh(z/b)^2=1$ for $z=h/2$ ($s=0$) and for general $z$ (general $t=s/L_0\tp{-1/2}$) yields
$L_0 = \sqrt{2(a^2-b^2)}$ and $r_0(t) = \sqrt{L_0^2\tp{t}^2 + b^2}$, respectively. Likwise, using the condition $H_0=0$ in the formula for the Gaussian curvature, eqn~(\ref{eqnKformula}), yields $K_0=-b^2/r_0^4$.  We expand 
\begin{eqnarray}
r(t) = r_0(t) +  r_1(t) + \hdots\\
H(t) = H_1(t) + \hdots\\
K(t) = -\dfrac{b^2}{r_0^4} +  K_1(t) + \hdots\\
\mu = \mu_0 +\mu_1 + \hdots
\end{eqnarray}
where $\epsilon=|L-L_0|/L_0\ll 1$, \tp{$r_0(t)$ is $O(\epsilon^0)$, $r_1(t)$ is $O(\epsilon)$, and so on.} 
Using eqn~(\ref{eqnnablaform}) in eqn~(\ref{shapeEq}) and working to order $\epsilon$, we have
\begin{equation}
\kappa (\Delta H_1 - 2K_0 H_1) = \mu_0 H_1
\label{evEq}
\end{equation}
with boundary conditions
\begin{equation}
H_1(t=\tp{-1/2}) = H_1 (t=\tp{1/2}) =  0.
\end{equation}
That is, the tension divided by $\kappa$ for a catenoidal membrane is an eigenvalue of the negative Jacobi operator, $-\mathcal{J}$, where
\begin{equation}
   \mathcal{J} = -\Delta + 2K.\label{jacobiop}
\end{equation}
Table~\ref{tab:jacobi} lists the first five eigenvalues of $-\mathcal{J}$ for a variety of $\bar{A}$; when $\bar{A} < 1$, they are all negative. The Jacobi operator arises most prominently in the formula for the second variation of area for a minimal surface~\cite{Simons1968,FomenkoTuzhilin1991}, thus connecting the tension of a catenoidal membrane to the stability of catenoidal soap films. 
Casting the problem in this form reveals that there are actually infinitely many modes of equilibrium solutions, each with two solution branches which meet when $h=h^*$ at a catenoid whose tension and force are negative. As mode number $n=1,2,3,\hdots$ increases, so does the number of oscillations in $r(z)$, as can be seen in Fig.~\ref{fig:catshapes}. The even symmetry of eqn~(\ref{evEq}) under $z\mapsto -z$ implies that the two branches are 
reflections of each other when $n$ is even, while the shapes are symmetric about the $z=0$ plane when $n$ is odd. 
These observations indicate that the determination of the tension of the catenoidal shapes is analogous to the Euler buckling problem of a solid thin rod under compression.~\cite{landau_lifshitz_elas} 
Fig.~\ref{fig:Fvh_smallA} illustrates the force as a function of extension for the first five modes. For all shapes, the force is negative (compressive), but the tension can be positive or negative. 
In section~\ref{Sec:stability}, we calculate the stability of the shapes, which is indicated in Fig.~\ref{fig:Fvh_smallA}.
Note that 
the higher order shapes are unstable.

\begin{table}
    \centering
    \begin{tabular}{|c|c|c|c|c|c|}
     \hline
     \multicolumn{6}{|c|}{Thick Catenoid Eigenvalues} \\
    \hline
     $\bar{A}$& $n=1$ & $n=2$ & $n=3$ & $n=4$ & $n=5$  \\ \hline
     0.2 & -243.6 & -948.8 & -2204 & -3920 & -6128 \\ \hline
     0.4 & -58.44 & -238.4 & -538.4 & -958.5 & -1499 \\ \hline
     0.6 & -24.03 & -101.0 & -229.3 & -409.0 & -640.1 \\ \hline
     0.8 & -11.79 & -52.49 & -120.2 & -215.0 & -336.9\\ \hline
     1 & -5.75 & -29.20 & -67.96 & -122.2 & -192.0 \\ \hline
     1.1 & -3.54 & -21.29 & -50.30 & -90.91 & -143.2 \\ \hline
     $\bar{A}_{\mathrm{max}}$ & 0 & -11.84 & -29.45 & -54.33 & -86.33 \\ \hline
    \end{tabular}
    \begin{tabular}{|c|c|c|c|c|c|c|}
     \hline
     \multicolumn{6}{|c|}{Thin Catenoid Eigenvalues} \\
    \hline
     $\bar{A}$& $n=1$ & $n=2$ & $n=3$ & $n=4$ & $n=5$  \\ \hline
     1.01 & 171.5 & -5.96 & -10.37 & -31.61 &-44.47\\ \hline
     1.05 & 22.95 & -6.50 & -14.72 & -34.38 & -54.64 \\ \hline
     1.1 & 8.45& -7.30 & -18.35 & -37.59 & -60.74 \\ \hline
     1.15 &  3.78 & -8.41 & -21.61 & -41.68 & -67.04 \\ \hline
     $\bar{A}_{\mathrm{max}}$ & 0 & -11.84 & -29.45 & -54.33 & -86.33 \\ \hline
    \end{tabular}
    \caption{The five largest numerically computed eigenvalues of the negative Jacobi operator $-\mathcal{J}$ [defined in eqn~(\ref{jacobiop})] for a selection of catenoids of dimensionless area $\bar{A}$. The eigenvalues of $-\mathcal{J}$ both govern the stability of catenoidal soap films and are the tensions $\mu^{(n)}/\kappa$ of fixed area catenoidal membranes. The eigenvalues for the thick catenoid are always negative, while the leading eigenvalue for the thin catenoid is always positive (not counting the critical case $\bar{A} = \bar{A}_{\mathrm{max}}$).}
    \label{tab:jacobi}
\end{table}

\subsubsection{Intermediate area: $1 < \bar{A} < \bar{A}_\mathrm{max}$}

\begin{figure*}[t]
\centering
\includegraphics[scale=0.119]{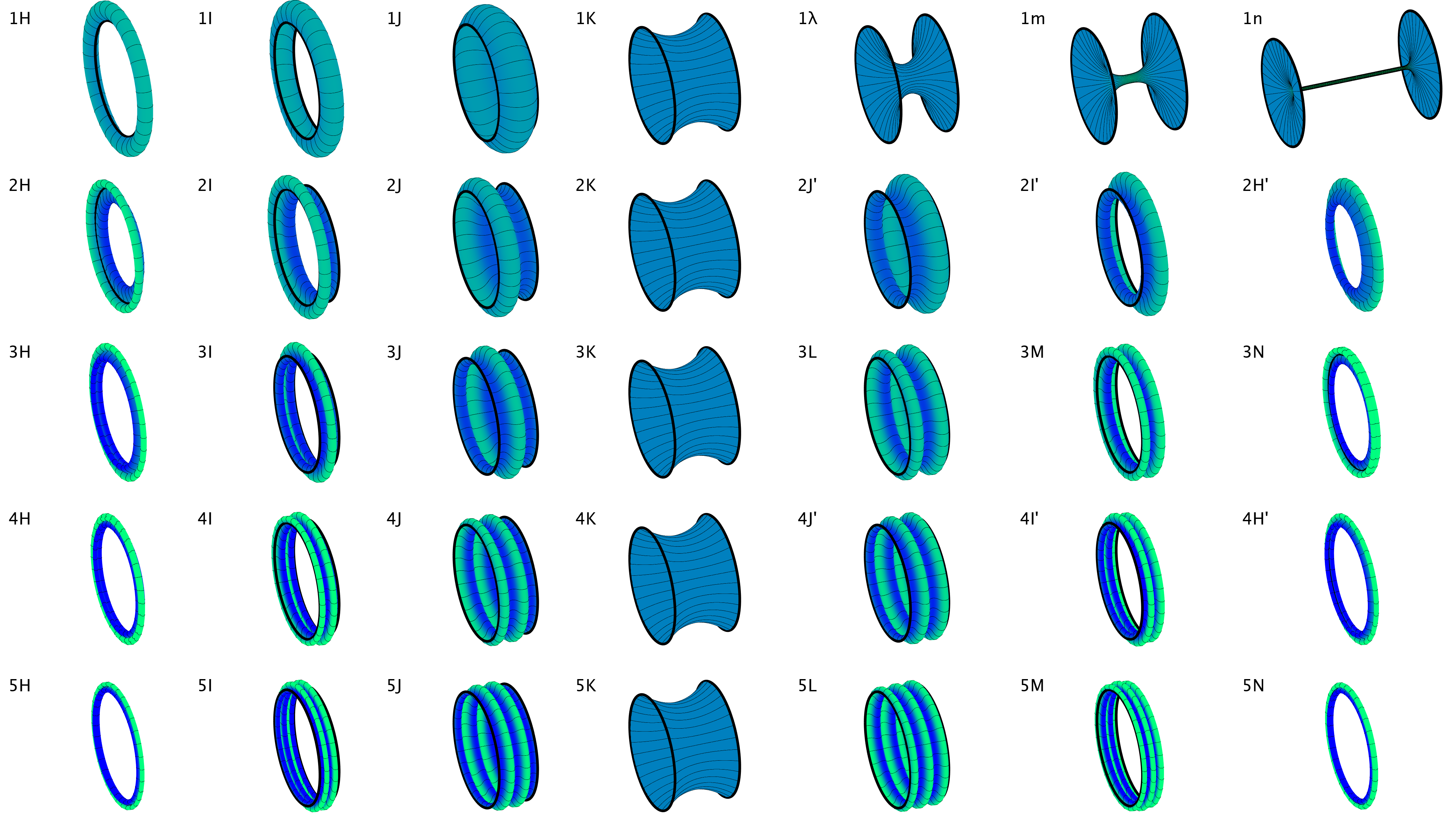}
\includegraphics[scale=0.25]{colorbar_v2.png}
\caption{ Representative shapes from the branches of the first five buckling modes connected to the thick catenoid for an axisymmetric membrane with $\bar\kappa = 0$ and dimensionless area $\bar{A} = 1.1$. Each row corresponds to one mode; upper (resp. lower) case letters denote a shape connected to the thick (resp. thin) catenoid ($\lambda$ is used in place of lower case L).
For $n=1$, the membrane can be continuously pulled from zero extension (H) into a thick catenoid (K), then compressed into a thin catenoid ($\lambda$). Pulling on the thin catenoid results in the emergence of a thin tether (m) connecting two catenoid-like ends; this tether has no maximal extension (n).
For $n>1$, the branch behavior depends on the parity of $n$: if $n$ is odd, the shapes are distinct, first traversing the branch with less negative force (H to J) to the thick catenoid (K), then the branch with more negative force (L to N). If $n$ is even, the shapes are mirror images of each other (denoted by a prime). Surfaces begin to self-intersect as $h$ approaches zero. Color indicates local mean curvature $Ha$. Black rings of radius $a$  have been added at $z=\pm h/2$ for visibility. 
}
\label{fig:catshapes2}
\end{figure*}

\begin{figure*}[t]
\centering
\includegraphics[scale=0.115]{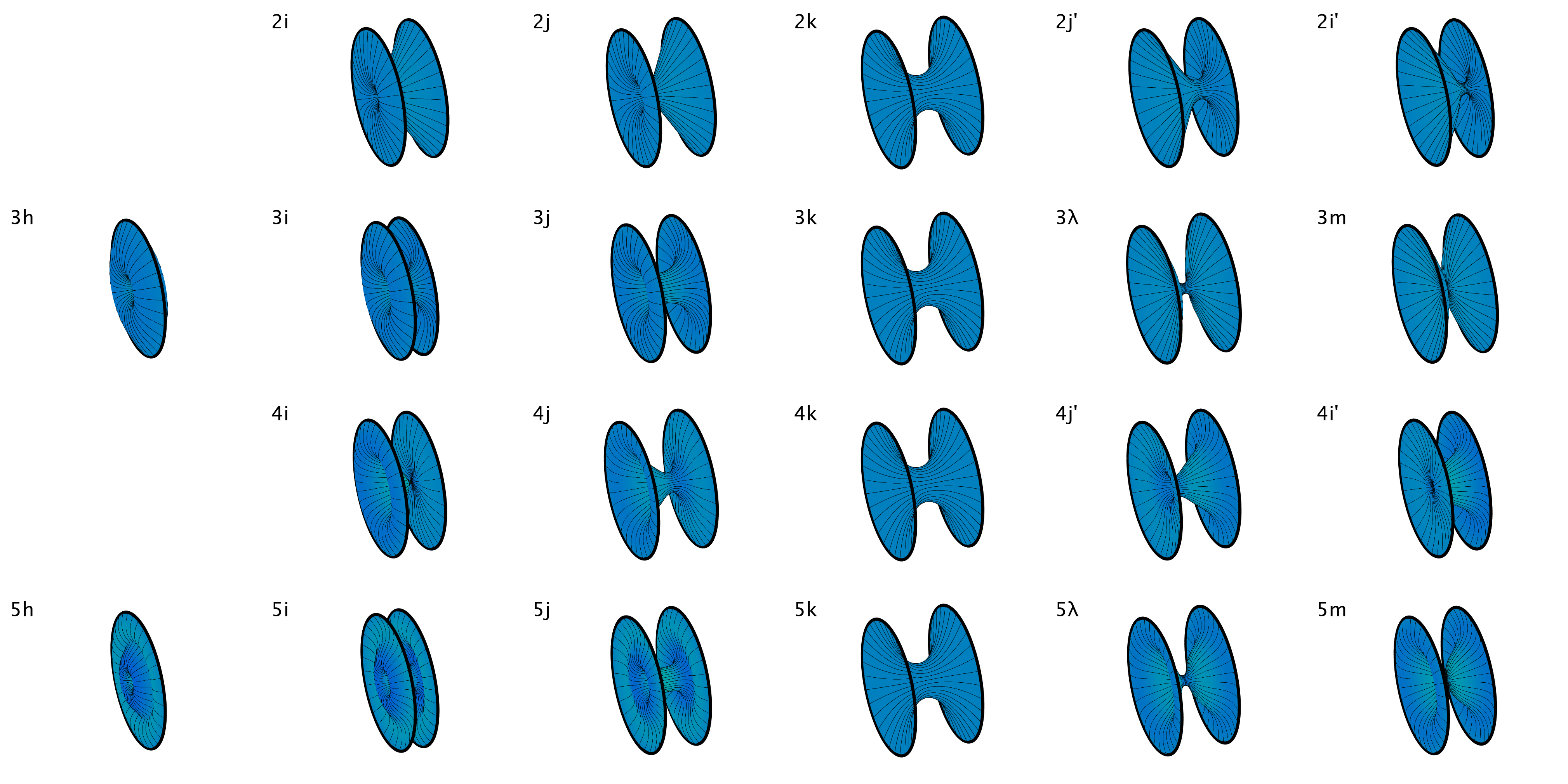}
\includegraphics[scale=0.25]{colorbar_v2.png}
\caption{ The second through fifth buckling modes connected to the thin catenoid for an axisymmetric membrane with $\bar\kappa = 0$ and dimensionless area $\bar{A}=1.1$. Each row contains shapes from two continuous branches that meet at the thin catenoid; the label $\lambda$ is used in place of lower case L. The branch behavior depends on the parity of $n$: if $n$ is odd, the shapes are distinct, first traversing the branch with more negative force with $h/a = 0$ (h), $h/a = 0.35$ (i), $h/a = 0.7$, and $h/a\approx 1.0428$ (k, the thin catenoid), followed by the branch with less negative force with an intermediate point ($\lambda$) and the point where the radius of the membrane collapses to zero (m). If $n$ is even, the shapes are mirror images of each other (denoted by a prime). Surfaces begin to self-intersect as $h$ approaches zero. Color indicates local mean curvature $Ha$.  Black rings of radius $a$  have been added at $z=\pm h/2$ for visibility. 
}
\label{fig:catshapes2b}
\end{figure*}

As $\bar{A}$ increases beyond unity, a thin catenoid emerges in addition to the existing thick catenoid as a possible solution. Repeating the perturbation argument in the previous section shows that each catenoid has infinitely many permissible tensions, and that locally there are two solution branches per tension that emanate from a catenoid. Qualitatively, the shapes from the branches corresponding to the thick catenoid resemble those from the $\bar{A} < 1$ case (Fig.~\ref{fig:catshapes2}), while those from the thin catenoid can look quite different, with necks that are comparatively much smaller (Fig.~\ref{fig:catshapes2b}). In fact, the necks of some of these shapes can even decrease to zero as they are compressed, effectively terminating the branch at some nonzero value of $h$. The thin solutions also have less energy than their thick counterparts and can have different stability properties.

\begin{figure}[t!]
\centering
\includegraphics[scale=0.2]{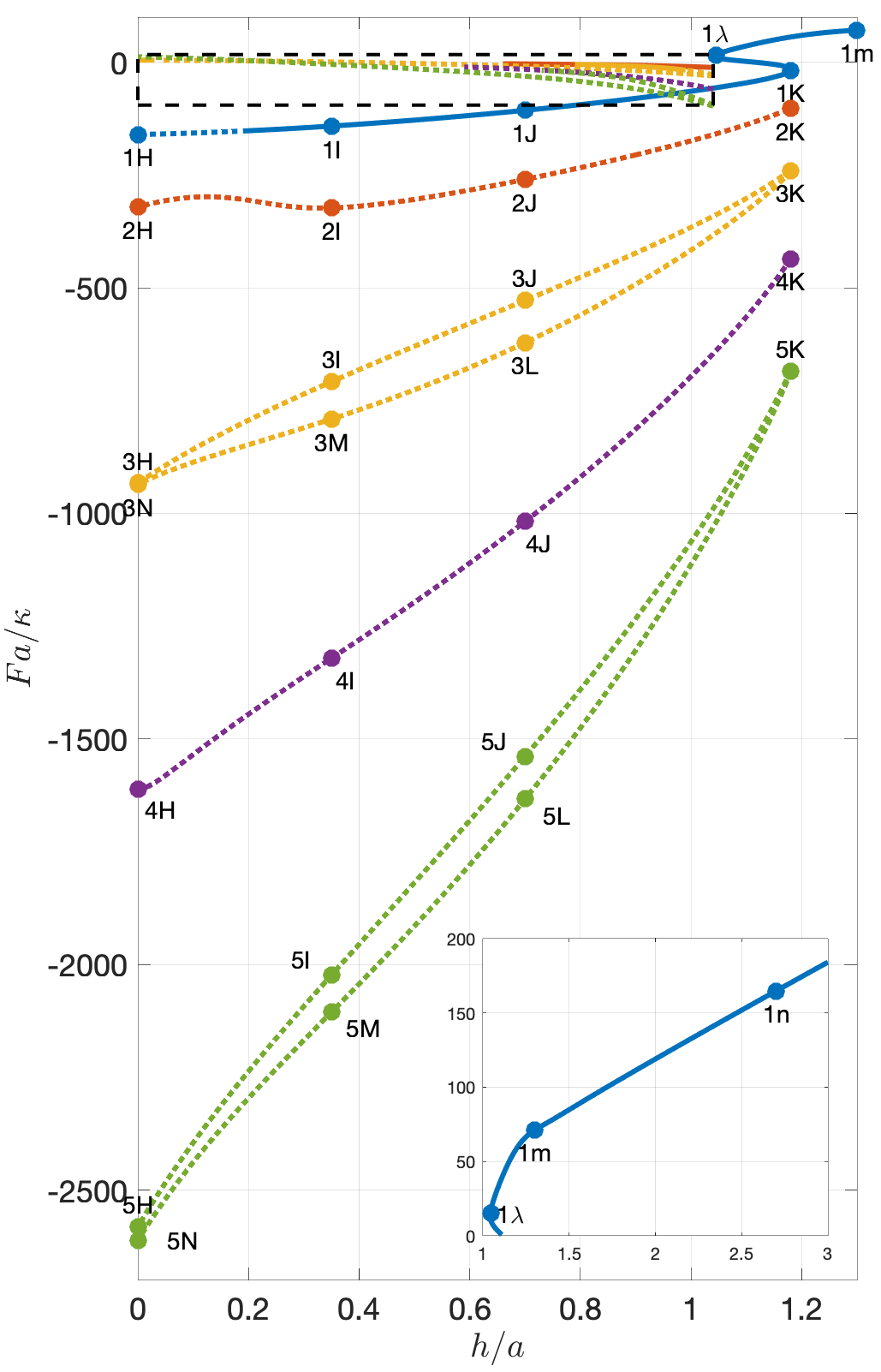}
\includegraphics[scale=0.2]{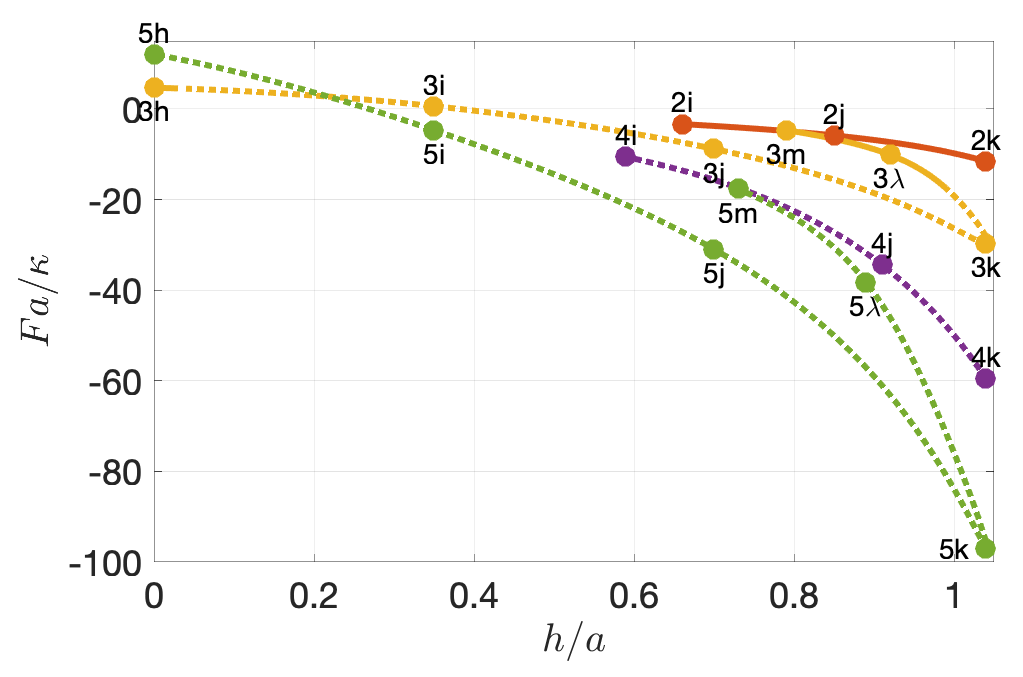}
\caption{ (a) Dimensionless force vs. extension for the first five bending modes of a membrane with $\bar{A}=1.1$ and $\bar\kappa = 0$. Labeled points correspond to the shapes in Figs.~\ref{fig:catshapes2} and~\ref{fig:catshapes2b}. Two possible catenoids with maximal extensions $h/a \approx 1.0428$ (thin) and $h/a \approx 1.1813$ (thick) can be formed with this area; locally, each catenoid has two solution branches per force eigenvalue.  Stable surfaces are indicated with a solid line while unstable surfaces are indicated with a dotted line. (Inset) The $n=1$ mode tether emerging from the thin catenoid has no maximal extension. As $h$ increases, the force becomes a linear function of extension.  (b) Larger version of the $n>1$ thin catenoid branches. On some of these branches, the membrane can have a minimum radius that goes to zero, terminating the branch at a nonzero extension. Note that in addition to the stable region near the shape 3$\lambda$,  there is a very small region of stability around the surface 3h.}
\label{fig:Fvh_medA}
\end{figure}

For this range of $\bar{A}$, significant differences between the $n=1$ mode and the $n>1$ modes develop. As Fig.~\ref{fig:Fvh_medA} shows, the $n=1$ mode is the only one where the thick and thin catenoids are connected by a path of equilibrium shapes. While the thick catenoid always has $F<0$ as before, the thin catenoid has $F>0$ for $n=1$. This property can be related back to the eigenvalues of $-\mathcal{J}$: since the thin catenoid is an unstable equilibrium of the area functional, its leading tension eigenvalue (and therefore its corresponding force) is positive (see Table~\ref{tab:jacobi}). A consequence of this sign difference is that one of the shapes on the connecting line between shapes $1$K and $1\lambda$ in Fig.~\ref{fig:Fvh_medA} is a free-floating surface with $F=0$. We find that the $n=1$ thin catenoid is a local minimum of extension rather than a maximum, leading to a possible hysteresis loop. (For $n>1$, the thin catenoid has negative tension and is a local maximum of extension just like the thick catenoid.) As the membrane is stretched beyond this thin catenoid, its neck can continue to decrease into a slender connecting tether (Figs.~\ref{fig:Fvh_medA} and~\ref{fig:catshapes2}, shapes 1m and 1n). Much like the Goldschmidt solution, and in contrast with the other branches, this tether solution has no maximal extension--the rings can be pulled arbitrarily far apart. Membrane tethers have been treated extensively elsewhere;~\cite{powers_huber_goldstein2002,DerenyiJulicherProst2002} here, we only recap their basic properties.


Past a certain extension, $H$ cannot be zero everywhere, and the membrane instead opts to form two partial catenoids at either end and connect them using the excess area. As extension keeps increasing, the connection becomes thin and cylindrical; this collapse of the neck is accompanied by  sharp increases in the force and tension (Fig.~\ref{fig:Fvh_medA} inset).
A crude approximation shows that the force increases linearly with extension while tension increases quadratically when $h\gg a$. Assume that a very thin tether of radius $b\ll a$ connects to the rings via two very flat catenoids. We approximate the area as $A\approx 2\pi a^2+2\pi b h$. Thus,
\begin{equation}
b=\frac{A-2\pi a^2}{2\pi h}.
\end{equation}
The energy is the sum of the bending energy and the tension times the area. We neglect the area of the catenoids since we assume they depend weakly on $h$ and $b$. Thus,
\begin{equation}
E\approx \frac{\pi\kappa h}{b}+2\pi h b\mu.
\end{equation}
The force is $F=\partial E/\partial h=\pi\kappa/b+2\pi b\mu$. Also, we must have $\partial E/\partial b=0$ (normal force balance), which implies $\mu=\kappa/2 b^2$. Putting it all together yields
\begin{eqnarray}
\mu&=&\frac{\kappa}{2}\frac{4\pi^2h^2}{(A-2\pi a^2)^2}\\
F&=&\frac{4\pi^2\kappa h}{A-2\pi a^2}.
\end{eqnarray}
It has been shown that, asymptotically, the ends are catenoids of neck radius $\sqrt{2\kappa/\mu}$, while the end of the tether profile is an exponentially decaying sinusoid with characteristic decay length $\sqrt{\kappa/\mu}$.~\cite{powers_huber_goldstein2002} We observe the same scalings for our shapes as $h \to \infty$. 




\subsubsection{Large area: $\bar{A} \geq \bar{A}_\mathrm{max}$}

For the critical value $\bar{A} = \bar{A}_\mathrm{max}$, the thin and thick catenoids are the same. The branches of each mode meet at this catenoid, which has extension $h_\mathrm{max}$. If the area is increased yet further, the membrane enters a regime where catenoids cannot be formed. Since there is no longer a  catenoid to serve as a base state around which to perturb, the linearization argument from the previous sections does not directly carry over. Regardless, there are some similarities with the previous cases.

\begin{figure}[h!]
\centering
\includegraphics[scale=0.2]{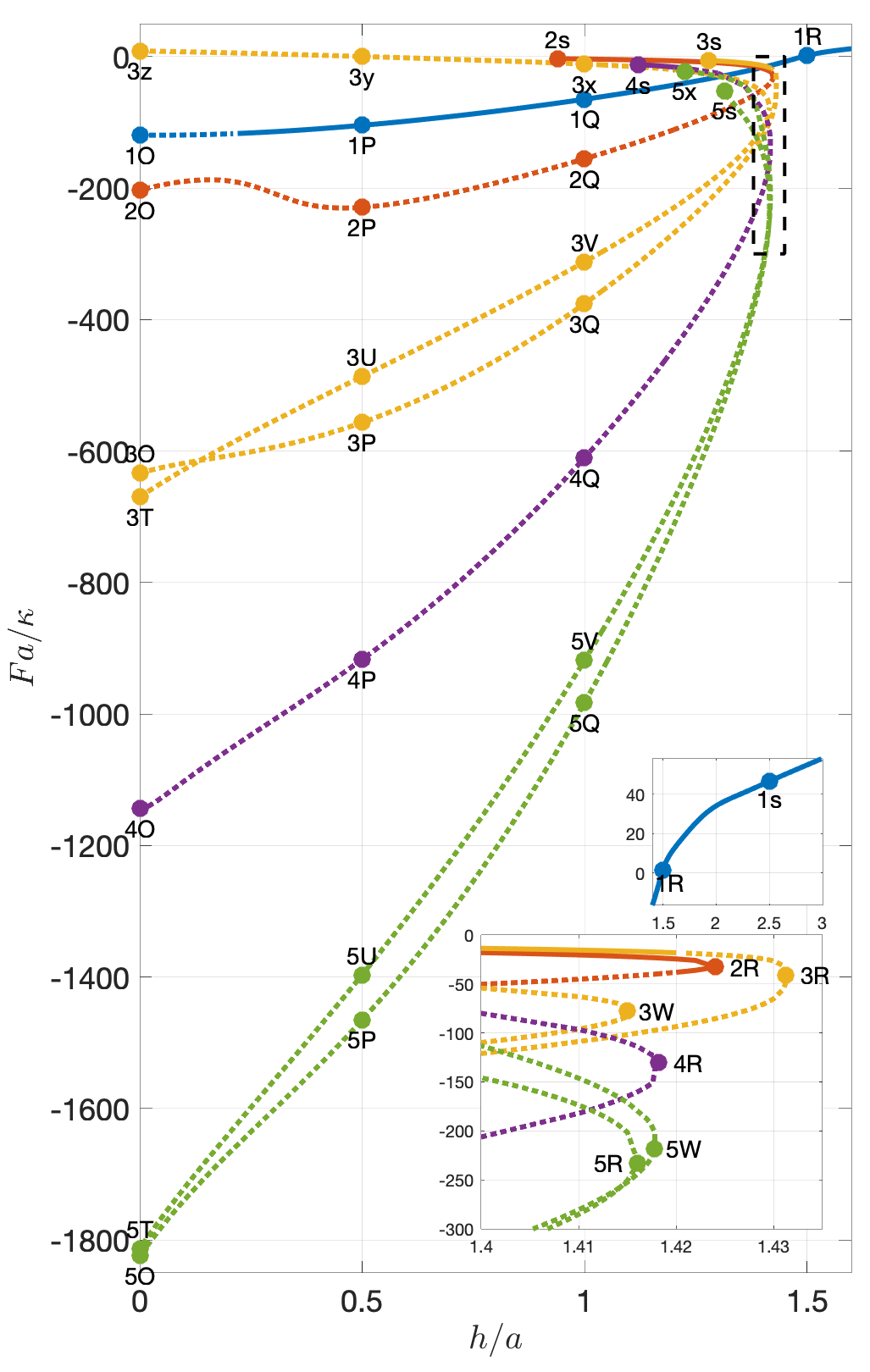}
\caption{Dimensionless force vs. extension for the first five bending modes of a membrane with $\bar{A}=1.3$ and $\bar\kappa = 0$.  Labeled points correspond to the shapes in Fig.~\ref{fig:catshapes3}. For even modes, the two branches are symmetric and hence their curves in the force vs. extension diagram overlap each other; for odd modes, the two curves are similar but not exact. Stable surfaces are indicated with a solid line while unstable surfaces are indicated with a dotted line. (Small inset) For $n=1$, the membrane can be pulled into a tether and has no maximal extension. (Large inset) Magnified version of the $n>1$ branches in the boxed region. For $n>1$, each branch has a different maximal extension at a shape that is not a catenoid. }
\label{fig:Fvh_largeA}
\end{figure}

First, when $n=1$, the membrane still develops a tether. However, since the area is too large for catenoids to form, there are no turning points where $dh/dF=0$. Instead, the force is a monotonically increasing function of extension and there is no hysteresis. Just as in the case of intermediate area,  there exists an equilibrium shape with $F=0$ on this branch. As shown in the smaller inset of Fig.~\ref{fig:Fvh_largeA}, the tether can still be arbitrarily long and thin, and the force continues to be a nearly linear function of extension in the large $h/a$ limit.

\begin{figure*}[ht!]
\centering
\includegraphics[scale=0.115]{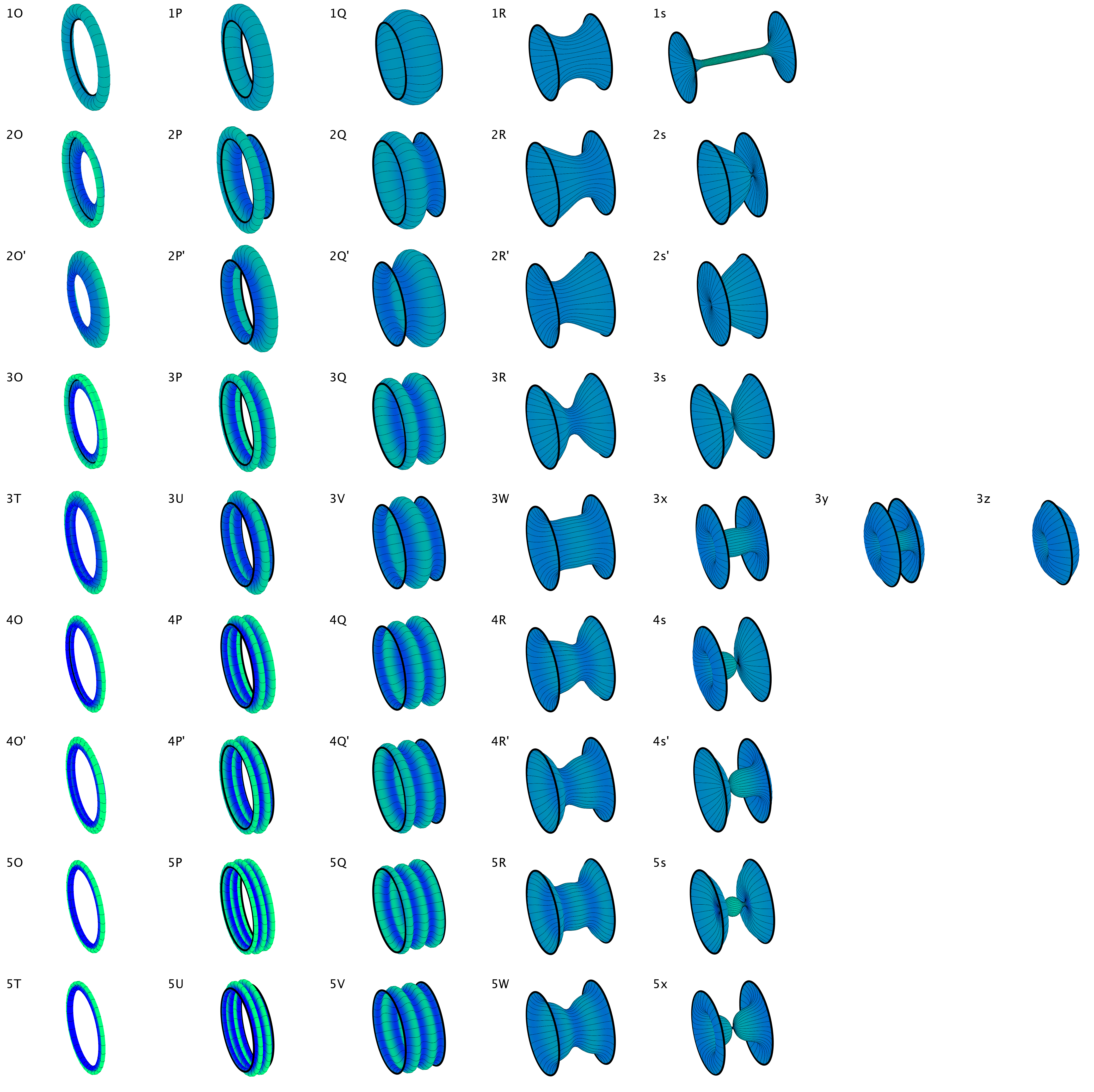}
\includegraphics[scale=0.25]{colorbar_v2.png}
\caption{The different branches of first five buckling modes of an axisymmetric membrane with $\bar\kappa = 0$ and dimensionless area $\bar{A}=1.3$. Each row contains shapes obtained from continuously traversing two branches of solutions of a single mode; upper (resp. lower) case letters denote a shape related to the thick (resp. thin) catenoid. For $n=1$, the membrane can be pulled directly into a thin tether with no maximal extension (s).
For $n>1$, the membrane can be pulled from zero extension through some shapes resembling thick catenoidal shapes (O to Q or T to V) to a non-catenoidal shape of maximal extension (R). Compressing this maximal shape can yield shapes that resemble thin catenoidal shapes (s or x to z). For some branches, the radius of the membrane approaches zero and thus the branch terminates. If $n$ is even, the two branches are mirror images of each other (denoted by a prime symbol). Color indicates local mean curvature $Ha$. Black rings of radius $a$  have been added at $z=\pm h/2$ for visibility. 
}
\label{fig:catshapes3}
\end{figure*}

For $n>1$, we still find that the branches have maximal extensions (large inset of Fig.~\ref{fig:Fvh_largeA}) that increase with area but are always finite. Unlike previous cases, different branches have different maximal extensions because the shape of maximal extension is no longer a catenoid. Some of these unusual shapes of maximal extension are shown in Fig.~\ref{fig:catshapes3}. Just as for smaller areas the catenoid served as a junction between two branches, so do these energy-minimizing shapes. A very notable difference, however, is that it is one branch of thin shapes and one branch of thick shapes that are joined, rather than two branches of the same kind. 
As before, some of the thin branches have a minimum radius that goes to zero at $h/a>0$, and the thin shapes have less energy than the corresponding thick shapes.


\subsection{Case of nonzero Gaussian curvature modulus}

For the case of $\bar\kappa \neq 0$, 
the aforementioned division into three area regimes still holds. Somewhat surprisingly, changing $\bar\kappa$ generally has a very weak effect on the membrane shapes, even if $\bar\kappa$ is comparable in magnitude to $\kappa$ (Fig.~\ref{fig:kbarshapes}a). The most prominent differences between the $\bar\kappa = 0$ and $\bar\kappa \neq 0$ cases are seen in the behavior of the force (Fig.~\ref{fig:kbarshapes}b) and tension.

\subsubsection{Small area: $\bar{A} \leq 1$}

\begin{figure}[ht!]
\centering
\includegraphics[scale=0.2]{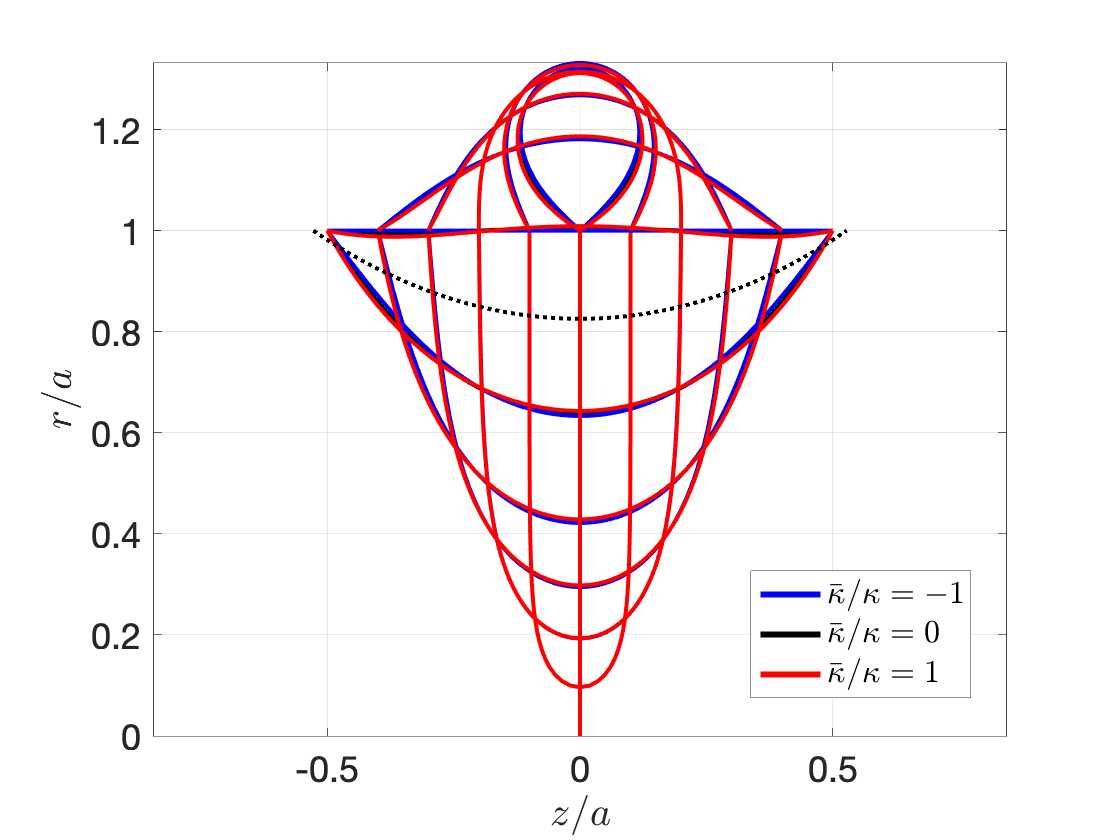}
\includegraphics[scale=0.2]{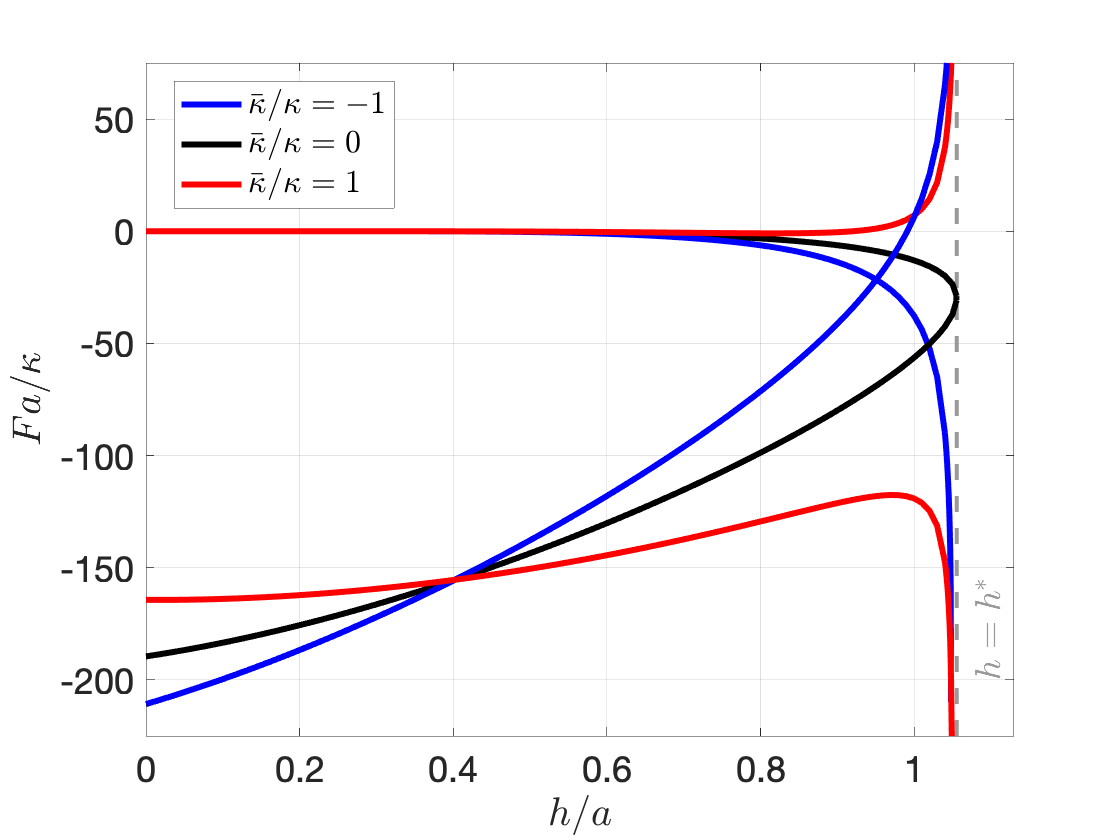}
\includegraphics[scale=0.2]{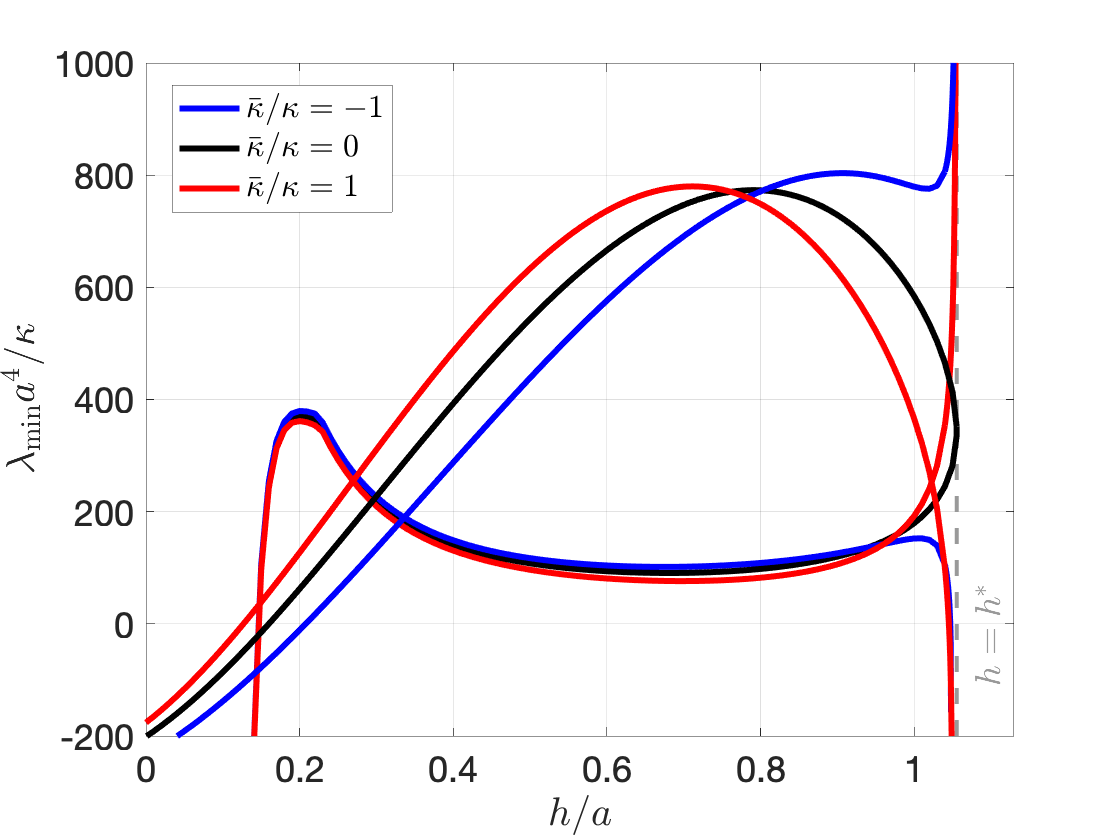}
\caption{ (a) Shape profiles for the $n=1$ buckling mode of an axisymmetric membrane with $\bar{A}=1$ and $\bar\kappa/\kappa=$ 0 (black), -1 (blue), and $1$ (red), showing very little difference in shape as $\bar\kappa$ is varied. The dotted line shows the catenoid, which is incompatible with boundary conditions when $\bar\kappa \neq 0$. (b) As the maximal extension $h=h^*$ (grey dashed line) is approached, the force diverges when $\bar\kappa \neq 0$. The direction of divergence depends on the sign of $\bar\kappa$ and the branch. (c) The minimal eigenvalue of the stability operator also exhibits a divergence at $h=h^*$ when $\bar\kappa \neq 0$. The sign of the divergence is the same as that of $F$. At $h=h^*$, when $\bar\kappa=0$, the minimal nonzero eigenvalue is given by $\mu^{(2)}(\mu^{(1)}-\mu^{(1)})/(2\kappa)$, where $\mu^{(n)}$ is the tension of the $n$th mode catenoid.}
\label{fig:kbarshapes}
\end{figure}

Being minimal surfaces, catenoids do not satisfy the no-torque boundary condition eqn~(\ref{notorque}) when $\bar\kappa \neq 0$. This leads to an apparent paradox: as we have seen, at the maximal extension, the catenoid is the unique axisymmetric surface, so the membrane must become more and more ``catenoid-like'' as it is pulled; yet, a true catenoid is unattainable. The resolution is that a nonzero Gaussian curvature modulus introduces singular behavior into the mean curvature. 

This singular behavior manifests itself in the force vs. extension plot as well. From the numerically calculated shapes, we observe that the force $F$ and tension $\mu$ diverge as $h\to h^*$ for all modes. Two branches are still present, but they are no longer connected. If $n$ is odd, one branch appears to go to positive infinity while the other goes to negative infinity (Fig.~\ref{fig:kbarshapes}b). If $n$ is even, both branches go to negative infinity (assuming $\bar\kappa > 0$). Changing the sign of $\bar\kappa$ reverses which branch goes to which infinity but $F$ and $\mu$ still blow up. However, the membrane profiles $r(z)$ qualitatively look very similar to the shapes from the $\bar\kappa = 0$ case: an infinite number of modes are still visible, each with two branches. 
The branches are not connected at $h=h^*$ due to the divergence of the force and tension at the maximal extension at $h=h^*$.

\tp{To determine how the tension diverges when $h$ approaches $h^*$, we use the observation from our numerical results that when $\bar{\kappa}\neq0$ and $h\to h^*$, the membrane shape is close to that of a catenoid except in thin boundary layers near the two edges. The width of each boundary layer is given by the natural length scale $\delta=\sqrt{\kappa/|\mu|}$ in the Euler-Lagrange eqn~(\ref{shapeEq}). In these boundary layers, the mean curvature $H$ changes rapidly, but $H$ and $K$ remain bounded. Therefore, the dominant balance for eqn~(\ref{shapeEq}) is}
\begin{equation}
\pm\delta^2\frac{d^2H}{ds^2} = H,
\label{blEq}
\end{equation}
\tp{with the}
sign on the left-hand side of eqn~(\ref{blEq}) matching the sign of $\mu$. 

\tp{The shape departs from a catenoidal shape because $h\neq h^*$, and because the no-torque boundary conditions [eqn~(\ref{notorque})] forbid $H=0$ near the edges with $\bar{\kappa}\neq0.$ In the following, we assume $\bar{\kappa}\ll\kappa$ to make analytical progress and because this limit is appropriate for colloidal membranes~\cite{Gibaud_etal2017,JiaZakharyDogicPelcovitsPowers2017}. For the catenoid of separation $h^*(A)$ connecting rings of radius $a$, the normal curvature of each edge is $k_\mathrm{n}=-b/a^2$. For $h$ near $h^*$, we have $k_\mathrm{n}=-b/a^2[1+O(\epsilon)]$, where $\epsilon=|L-L_0|/L_0$ was introduced in sec~\ref{smallareasec}. Thus, to leading order in the small quantities $\epsilon$ and $\bar{\kappa}/\kappa$, the no-torque condition is} 
\begin{equation}
\left. 2\kappa H- \bar\kappa \frac{b}{a^2}\right|_{s=0,L} = 0,
\end{equation}
There are two cases to consider: $\mu$ positive or negative. 
Once $H$ is known, \tp{we calculate the shape from the definition of the mean curvature and use the constraint of constant area to find the tension. In the following we focus on the case of positive tension; the case of negative tension is discussed in the ESI$^\dag$.}

When $\mu > 0$, the solution \tp{for the mean curvature when $\delta\ll1$ and to leading order in $\bar\kappa/\kappa$} is
\begin{equation}
H = \frac{\bar\kappa b}{2\kappa a^2} \frac{\cosh\left[(s-{L}/{2})/\delta\right]}{ \cosh[L/(2\delta)]}.\label{Hvs}
\end{equation}
\tp{Note that} $H$ is 
\tp{exponentially small}  
except near the endpoints, where it exhibits boundary layers of width $O (\delta)$.
\tp{Given the mean curvature, we solve for the shape $r=r_0+r_1$, where $r_0=\sqrt{t^2L_0^2+b^2}$ is the catenoid shape in terms of the dimensionless coordinate $t=s/L_0-1/2$ as in sec~\ref{smallareasec}, and $r_1$ is a perturbation that vanishes when $\epsilon\to0$.} As is traditional, we divide the domain into  inner 
\tp{regions near edges} and  \tp{an} outer \tp{region} 
where \tp{the} mean curvature \tp{is} approximately zero.  
\tp{Then we} approximately solve for $r_1$ in each region, and match the two solutions to generate a composite solution.

\tp{First consider the outer region where $H$ is exponentially small. Using eqn~(\ref{Hitopsirs}), $r_s=\cos\psi$, and $L=L_0(1+\epsilon)$ to expand $H=0$ to first order in $r_1=r^\mathrm{out}$ and $\epsilon$, we find}
\begin{equation}
\frac{r_0}{bL_0^2} \frac{d^2 r^\mathrm{out}}{dt^2}  + \frac{\tp{2t}}{br_0} \frac{dr^\mathrm{out}}{dt} + \frac{b}{r_0^3}r^\mathrm{out}-2\frac{\epsilon}{b}=0,
\label{r1Eq}
\end{equation}
\tp{which has solution}
\begin{eqnarray}
    r^\mathrm{out}
&=&  \frac{\epsilon L_0^2 \tp{t}^2+C_0}{r_0(t)}\\
&=&\frac{\epsilon  (s-L_0/2)^2+C_0}{r_0(s)}. 
\end{eqnarray}
Note that we used reflection symmetry about $t=\tp{0}$ to determine the integration constant that multiplies the solution $t/r_0(t)$.


\tp{Next, consider the inner regions,}  such as the region near the  
endpoint $t=\tp{-1/2}$. \tp{Since $(t+1/2)L=s\lesssim\delta\ll1$ in this region, we may take $r_0\approx a$ and $\sqrt{1-(dr_0/ds)^2}\approx b/a$. Furthermore, the second derivative term dominates the mean curvature, and we may write the equation for mean curvature in terms of $r_1=r^\mathrm{in}$ as}
\begin{equation}
 \frac{a}{b}\frac{d^2 r^\mathrm{in}}{ds^2}=   \frac{\bar{\kappa}b}{\kappa a^2}\exp(-s/\delta),
\end{equation}
\tp{where on the right-hand side we have written $H$ for small $\delta/L$.}
The solution \tp{in the inner region near $s=0$} has the form
\begin{equation}
\tp{r^\mathrm{in} = \frac{\bar\kappa b^2}{\mu a^3}  \exp(-s/\delta) + C_1s - \frac{\bar\kappa b^2}{\mu a^3}}
\end{equation}
where we have made use of the boundary condition $r\tp{^\mathrm{in}} = 0$ at the endpoint \tp{$s=0$}. It remains to solve for the constants $C_0$ and $C_1$ from matching. Since 
\tp{the linear term of the inner solution cannot match with the outer solution, $C_1$} is zero. As for $C_0$, we calculate the overlapping part and find
\begin{equation}
\lim_{s\to 0} r^\mathrm{out} = \frac{\epsilon L_0^2/4+C_0}{a}= -\frac{\bar\kappa b^2}{\mu a^3} = \lim_{s/\delta \to \infty } r^\mathrm{in}
\end{equation}
The uniformly accurate composite approximation is then given by the sum of the inner and outer solutions minus the overlapping part. \tp{We use reflection symmetry about $s=L_0/2$ to get the correct expression near $s=L_0$:}
\begin{equation}
    r_1=\frac{\bar{\kappa}b^2}{\mu a^3}\left[\mathrm{e}^{-s/\delta}+\mathrm{e}^{-(L_0-s)/\delta}-\frac{a}{r_0(s)}\right]+\frac{\epsilon s(s-L_0)}{r_0(s)}.\label{r1comp}
\end{equation}
\tp{The excellent agreement between} 
the numerically computed solution 
\tp{and} the approximation eqn~(\ref{r1comp}) \tp{for small $\epsilon$} is shown in the ESI$^\dag$. Using the fact that \tp{the area constraint implies that} the integral of $r_1(s)$ 
\tp{vanishes to leading order,} eqn~(\ref{r1comp}) implies the scaling law
\begin{equation}
\mu \propto\frac{\bar\kappa}{a^2}\epsilon^{\tp{-1}}
\end{equation}
as $h\to h^*$. \tp{Using $dz/ds^2+dr/ds^2=1$ and the area constraint leads to the relation $\mu\propto[(h^*-h)/h^*]^{-2/3}$ for $h$ near $h^*$.} To leading order, $F\sim 2\pi \mu b$ in this limit as before. These relations are independent of mode (that is, the tension vs. extension or force vs. extension curves for each mode all collapse in the $h\to h^*$ limit).

If instead we have $\mu <0$, the mean curvature to leading order is
\begin{equation}
H = \frac{\bar\kappa b}{2\kappa a^2} \frac{\cos \left[(s-{L}/{2})/\delta\right]}{ \cos [L/(2\delta) ]}.
\label{negmuH}
\end{equation}
Thus in this limit the mean curvature oscillates rapidly but converges weakly to zero.
Note that eqn~(\ref{negmuH}) is a poor approximation when $L_0\sqrt{|\mu| /\kappa }= k\pi$, where $k$ is an odd integer. The linearized membrane shape equation has an infinite sequence of eigenvalues when $\mu < 0$, and by the Fredholm Alternative, we cannot expect our inhomogeneous problem to be solvable at these points. Eqn~(\ref{blEq}) with Dirichlet boundary conditions has eigenvalues at $\delta = L_0/k\pi$, which leads to a poor approximation whenever $\delta$ approaches these values. Refining the approximation with higher order terms  alleviates this issue but for simplicity of presentation we will only consider leading order terms here.

To approximately solve eqn~(\ref{r1Eq}), we  split the equation into two parts: an ``oscillatory'' part, $r_1^{(\mathrm{osc})}$, that solves the inhomogeneous equation with the oscillatory $H$ forcing and a ``remaining'' part, $r_1^{(\mathrm{rem})}$, that solves the equation with the remaining $\epsilon$ term. The boundary conditions for the ``remaining'' part will be chosen so that the sum adds up to zero at the boundaries. Thus, we are solving
\begin{equation}
\frac{r_0}{bL_0^2} \frac{d^2 r_1^{(\mathrm{osc})}}{dt}+\frac{\tp{2t}}{br_0} \frac{dr_1^{(\mathrm{osc})}}{dt} + \frac{b}{r_0^3} r_1^{(\mathrm{osc})} = \frac{\bar\kappa b}{2\kappa a^2} \frac{\cos[(s-L_0/2)/\delta]}{\cos[L_0/(2\delta)]}
\end{equation}
and
\begin{equation}
\frac{r_0}{bL_0^2} \frac{d^2 r_1^{(\mathrm{rem})}}{dt}+\frac{\tp{2t}}{br_0} \frac{dr_1^{(\mathrm{rem})}}{dt} + \frac{b}{r_0^3} r_1^{(\mathrm{rem})} = -\frac{2\epsilon}{b}
\end{equation}
subject to
\begin{equation}
r_1^{(\mathrm{rem})}(t=\tp{-1/2},\tp{1/2}) = -r_1^{(\mathrm{osc})}(t=\tp{-1/2},\tp{1/2}).
\end{equation}
For $r_1^{(osc)}$, we use the WKB approximation and find
\begin{equation}
r_1^{(\mathrm{osc})} = -\frac{\bar\kappa b^2}{\mu a^2 r_0} \frac{\cos \left[(s-{L_0}/{2})/\delta\right]}{\cos[L_0/(2\delta)]},
\label{r1osc}
\end{equation}
and consequently,
\begin{equation}
r_1^{(\mathrm{rem})} = \frac{\bar\kappa b^2}{\mu a^3} + 2\epsilon \left[a -r_0 -\frac{L_0}{2}\tanh^{-1} \frac{L_0}{2a}+\left(s-\frac{L_0}{2}\right)\tanh^{-1}\frac{s-L_0/2}{r_0} \right].
\label{r1rem}
\end{equation}
Our leading order perturbation is the sum of eqn~(\ref{r1osc}) and~(\ref{r1rem}) (again, this expression doesn't apply near eigenvalues) and is plotted in the ESI$^\dag$. Upon applying the area constraint, we again find the general approximation $\mu \sim \bar\kappa \epsilon^{-1}/a^2$ for the negative tension branch.

While in the case of zero Gaussian curvature modulus all shapes with $\bar{A} < 1$ required compressive external forces, the divergence of $F$ for $\bar\kappa \neq 0$ will make the force for one branch positive for $n$ odd and $h$ near $h^*$ (the sign of $\bar\kappa$ determines which branch). This implies the existence of an equilibrium shape with $F=0$. In short: nonzero Gaussian curvature modulus is necessary in order to have a free-standing shape with $\bar{A} < 1$, and this shape is very nearly a catenoid.

\subsubsection{Intermediate area: $1<\bar{A} < \bar{A}_\mathrm{max}$}

The arguments in the previous case can be generalized in a straightforward manner to show that tension and force diverge when either catenoid is approached. Thus, while it was possible to continuously deform a thick catenoid into a thin catenoid and into a tether when $\bar\kappa$ was zero, this is prohibited when $\bar\kappa \neq 0$ due to the divergences near each catenoid. The scalings in the previous section are seen to hold near each catenoid.

For $n=1$, tethers are still observed; qualitatively they resemble the tethers from the $\bar\kappa = 0$ case in shape. It is interesting to note that if $\bar\kappa > 0$, the force is no longer a monotonically increasing function of extension. Instead, the formation of the tether coincides with a drop in $F$, after which $F$ returns to monotonically increasing as $h$ increases. This kind of behavior has been observed in other works~\cite{powers_huber_goldstein2002}. 

\subsubsection{Large area: $\bar{A} \geq \bar{A}_\mathrm{max}$}

Analogous to the $\bar\kappa = 0$ case, the tether is still a valid $n=1$ solution. The properties described in the previous section are observed to hold here as well.

Higher order modes still have finite extensions. However, since there is no reason for the shapes to become catenoid-like as they are pulled, the tension and force do not blow up as the maximal extension is reached, unlike the $\bar{A} < \bar{A}_\mathrm{max}$ case. The maximal extension is observed to depend on $\bar\kappa$, albeit weakly.

\subsection{Special isolated shapes}


Here, we determine the conditions under which the membrane assumes a spherical, cylindrical, or Willmore toroidal shape. These are simple analytical limits of the $n=1$ mode described above which can easily be verified to satisfy the membrane shape equation eqn~(\ref{shapeEq}). Since we assume a certain shape profile, the area and extension need to be chosen consistently; consequently, these are isolated solutions that do not persist when the extension is varied. Previous work~\cite{Tu2010,Tu2011} has ruled out the existence of such shapes in force-free settings, but here we demonstrate they exist if the correct external forces are applied \tp{and the ratio $\bar{\kappa}/\kappa$ is tuned to a special value. For spheres and cylinders, this special value is negative, and therefore we do not find sections of spheres or cylinders in our numerical calculations, which have $\bar{\kappa}>0$.}

\subsubsection{Spheres}

Since spheres are easily seen to solve the membrane shape equation eqn~(\ref{shapeEq}) with $\mu  = 0$, we expect that if the boundary conditions allow for it, the membrane will assume this configuration. Our sphere will have caps missing due to boundary conditions; regardless, such a shape must satisfy $H = \tp{-}1/b$ and $K = H^2$, where $b>a$ is the radius of the sphere. \tp{Using $k_\mathrm{n}=-1/b$ for the normal curvature of a latitude of a sphere of radius $b$ in} the no-torque boundary condition eqn~(\ref{notorque}) allows one to deduce the requirement $\bar\kappa = \tp{-}2\kappa$ for sphere formation. The extension $h$ at which we have a sphere is given by the Pythagorean theorem, $h = 2\sqrt{b^2 -a^2}$,
while the area must also be consistently chosen, $A = 4\pi b^2 \sqrt{1- (a/b)^2}$.
Since the tension of a spherical membrane is always zero, Eqn~(\ref{forceEq}) confirms that the axial force is also zero. However, since we have fixed $r=a$ at the boundaries, external radial forces act at the edge of the membrane, and there is no contradiction with the nonexistence theorem for portions of membrane spheres with edges in the absence of external forces~\cite{Tu2010,Tu2011}. 

\subsubsection{Cylinders}

A cylinder has $2H=\tp{-}1/a$, $K = 0$, \tp{and $k_\mathrm{n}=-1/a$ for a latitude}; plugging these into eqn~(\ref{notorque}) yields the necessary condition $\bar\kappa =\tp{-}\kappa$ to satisfy the no-torque boundary condition. Then, if the area and extension satisfy $A = 2\pi a h$, we will have a cylinder. As can be seen from the membrane shape equation~(\ref{shapeEq}), the tension of a cylinder is always $\mu  =\kappa/2a^2$; consequently, the force is $F=\tp{2\pi \kappa/a}$. An example cylinder can be seen in Fig.~\ref{fig:kbarshapes}a when $h/a=1$.

\subsubsection{Willmore tori}

We parameterize the torus by $\mathbf{X}(s_1,s_2)=R_1\hat{\boldsymbol\rho}+R_2(-\cos s_2 \hat{\boldsymbol\rho}+\sin s_2\hat{\mathbf z})$, where $\hat{\boldsymbol\rho}=\hat{\mathbf{x}}\cos(s_1/R_1)+\hat{\mathbf{y}}\sin(s_1/R_1).$ Note that $s_1$ is a length, and $s_2$ is an angle. Using the formulas from Appendix~\ref{dgappx} (see also Willmore's textbook~\cite{Willmore1993}), we find
\begin{eqnarray}
\mathrm{d}A&=&R_2\left[1-({R_2}\cos s_2)/R_1\right]\mathrm{d}s_1\mathrm{d}s_2\\
H&=&-\frac{1-2(R_2\cos s_2)/R_1}{2R_2[1-(R_1\cos s_2)/R_2]}\\
K&=&-\frac{1}{R_1 R_2}\frac{\cos s_2}{1-(R_2\cos s_2)/R_1}
\end{eqnarray}
and 
\begin{equation}
    \Delta H+2H(H^2-K)=\frac{1-2R_2^2/R_1^2}{4R_2^3\left[1-({R_2}\cos s_2)/R_1\right]^3}.
\end{equation}
Thus we see that the torus satisfies the Euler-Lagrange equation~(\ref{shapeEq}) with zero tension if $R_2/R_1=1/\sqrt{2}$.

Next, we construct two different axisymmetric surfaces by cutting the torus along two circular latitudes at $s_2=\pm s_0$, as in Fig~.~\ref{fig:torus}, where the the ``outer" surface is green and the ``inner" surface is blue. The circular edges are the suspending rings of radius $a$, where for the Willmore torus we have $a=R_1-R_2 \cos s_0=R_1[1-(\cos s_0)/\sqrt{2}]$. As in the case of the spherical and cylindrical sections discussed in the preceding subsections, the condition of zero torque, eqn~(\ref{notorque}), leads to a condition on $\bar{\kappa}/\kappa$. However, unlike the sphere and the cylinder, this condition depends on where we cut the surface. For either the inner or the outer shape, the normal curvature is given by $k_\mathrm{n}=-\cos s_0/a$. Combining the no-torque condition~(\ref{notorque}) with the formula for the mean curvature for the Willmore torus yields 
\begin{equation}
    R_1=a\frac{2\kappa+\bar{\kappa}}{\kappa+\bar{\kappa}}\label{kbkappacond}
\end{equation}
for both the inner and outer surface. Writing $a$ in terms of $R_1$ and $s_0$, eqn~(\ref{kbkappacond}) becomes
\begin{equation}
\cos s_0=\frac{\sqrt{2}\kappa}{2\kappa+\bar{\kappa}}\label{s0eqn}.
\end{equation}
In other words, the value of $\bar{\kappa}/\kappa$ determines what portion of the torus satisfies the equilibrium conditions. For example, if $\bar{\kappa}=0$, then $s_0=\pi/4$, as in Fig.~\ref{fig:torus}, where the edges are curves with zero mean curvature. Note that  eqn~(\ref{s0eqn}) has a real solution only for $\bar{\kappa}>-2+\sqrt{2}$.
Once $s_0$ is found, then $R_1$ is determined by the area constraint. 
For example, when $\bar{\kappa}=0$, we find that $R1=2a$ and the reduced area of the inner surface is $\bar{A}=\sqrt{2}(\pi-2)\approx1.6145$, and the reduced area of the outer surface is $\bar{A}=\sqrt{2}(3\pi +2)\approx 16.16$. Oddly, the ratio of the areas of the outer and inner Willmore surface portions is very close to ten: $(3\pi+2)/(\pi-2)\approx10.008$. 

The force is conveniently found by using the cylindrical coordinate $r=R_1\pm\sqrt{R_1^2/2-z^2}$ and eqn~(\ref{forceEq}) with $\mu=0$, which yields
\begin{equation}
    F=\pm2\pi\sqrt{2} \kappa/R_1,
\end{equation}
with the plus sign for the inner surface and the minus sign for the outer surface. Again, for $\bar{\kappa}=0$, $F=\pm \pi\sqrt{2}\kappa/a$.

\subsection{Stability}\label{Sec:stability}



 \label{secondvarappendix}
 
 In order to analyze the stability of our surfaces, we calculate the second variation of eqn~(\ref{EEq}) \cite{zhong-can_helfrich1989,CapovillaGuvenSantiago2003} in the coordinate system shown in Fig.~\ref{fig:setup}. While the formula that appears in these references was derived for closed surfaces and hence does not include a Gaussian curvature term, it is straightforward to compute the variation of this term, which only appears at the boundary thanks to the Gauss-Bonnet theorem. Assuming an axisymmetric perturbation $u(s)$ to the surface in the normal direction, and using the formulas in Appendix~\ref{dgappx},  the second variation is $\delta^{(2)}E=\int dA u \mathcal{L}u$, where  
  \begin{equation}
      \mathcal{L} u = G_2(s) \Delta^2u + G_1(s) \Delta u + G_0(s) u,
  \end{equation}
\begin{equation}
G_2(s) = \frac{\kappa}{2},
\end{equation}
\begin{equation}
    G_1(s) = \kappa(H^2-2K) - 2\kappa H\psi_s - \frac{\mu}{2},
\end{equation}
and
\begin{align}
    G_0(s) &= \kappa [2(K- 4H^2)( K- H^2) -\psi_s H_{ss} -\frac{r_s\sin\psi}{r^2}H_s -H\Delta H+H_s^2]\nonumber\\
    &+\mu K - \frac{\kappa}{r} (r_s H \psi_s)_s +  \frac{\kappa}{r} \left(\frac{Hr_s\sin\psi}{2r}\right)_s.
\end{align}
The reader is cautioned that there is a discrepancy regarding the formula for $G_0$ in the references we have cited. Here, we calculate $G_0$ using the formula that appears in ref~\cite{CapovillaGuvenSantiago2003}, which claims to have corrected the one that appears in ref~\cite{zhong-can_helfrich1989}. Numerical tests indicate that the discrepancy does not meaningfully affect any of the results for our system. 

It is convenient to use integration by parts and the fact that $G_2$ is constant to write $\mathcal{L}$ in the symmetric form
\begin{equation}
    \mathcal{L}u=\frac{1}{r}\left[rF_2(s)u_{ss}\right]_{ss}+\frac{1}{r}\left[rF_1(s)u_s\right]_s+\frac{1}{r}\left[rF_0(s)u\right],
\end{equation}
with
\begin{equation}
    F_2(s) = G_2,
\end{equation}
\begin{equation}
    F_1(s) = G_1(s) - \frac{G_2}{r}\left(\frac{r_s^2}{r}-r_{ss}\right)
\end{equation}
and
\begin{equation}
    F_0(s) = G_0(s) + \frac{1}{2}\Delta G_1(s)
\end{equation}
The boundary conditions associated with this variation are 
\begin{equation}
    \left. u\right|_{s=0} = \left. u\right|_{s=L} = 0
    \label{stabbc1}
\end{equation}
and
\begin{equation}
   \left. \kappa\Delta u+\bar\kappa \frac{\cos\psi}{r}u_s \right|_{s=0} =\left. \kappa\Delta u +\bar\kappa \frac{\cos\psi}{r}u_s \right|_{s=L}= 0.
   \label{stabbc2}
\end{equation}
Note that $\mathcal{L}$ with these boundary conditions is self-adjoint.
We must also ensure that the perturbation does not change the area to first order, which yields an additional orthogonality constraint
\begin{equation}
    \int dA 2Hu = 0.
    \label{orthogconstraint}
\end{equation}

We thus need to solve a constrained eigenvalue problem~\cite{Golub1973,Nurse_etal2015} of the form
\begin{equation}
    \mathcal{L}u + 2H p = \lambda u 
    \label{stabEVeq}
\end{equation}
for eigenvalue $\lambda$ where $p$ is a Lagrange multiplier that enforces the orthogonality constraint eqn (\ref{orthogconstraint}). Taking the inner product of both sides of this equation with $2H$ reveals
\begin{equation}
    p = -\frac{\int dA 2H \mathcal{L} u}{ \int dA (2H)^2},
\end{equation}
which upon substitution converts eqn (\ref{stabEVeq}) into an unconstrained eigenvalue problem:
\begin{equation}
    \mathcal{P}\mathcal{L}u = \lambda u,
\end{equation}
where $\mathcal{P}$ is a projection onto the subspace $(2H)^\perp$. From this formulation, it is clear that $u=2H$ is an eigenfunction with eigenvalue zero; the remaining eigenfunctions are orthogonal to $2H$, and $p$ is zero for these eigenfunctions. Since we are interested in the smallest nonzero eigenvalue, we first discretize the operator $\mathcal{P}^\dagger \mathcal{L}\mathcal{P}$ (which is equivalent to discretizing $\mathcal{P}\mathcal{L}$ but has the advantage of being symmetric) using central finite differences, taking care to satisfy eqn (\ref{stabbc1}) and eqn (\ref{stabbc2}) at the interval endpoints, and solve a standard matrix eigenvalue problem. The smallest nonzero eigenvalue of this matrix and corresponding eigenvector are used as initial guesses for MATLAB's \texttt{bvp4c} in a routine that mirrors the one described in Section 2.2. As before, we consider separately the two cases  $\bar\kappa = 0$ and $\bar\kappa \neq 0$.

\subsubsection{Case of zero Gaussian curvature modulus}

While analysis of the general expression for $\mathcal{L}$ requires a numerical routine, the stability of the catenoids can be determined more readily by exploiting the intimate connection between Willmore stability and area stability for minimal surfaces. For the $n$th mode catenoid with tension $\mu^{(n)}$, the stability operator simplifies to 
\begin{equation}
\mathcal{L}=\frac{\kappa}{2}\mathcal{J}^2 + \frac{\mu^{(n)}}{2}\mathcal{J}.
\end{equation}
Recall that for catenoidal membranes, the allowable tensions $\mu^{(n)}/\kappa$ are precisely the eigenvalues of 
the negative of the Jacobi operator $\mathcal{J}$ [eqn~\ref{jacobiop}]. Furthermore, note that any eigenfunction of $\mathcal{J}$ is an eigenfunction of $\mathcal{L}$ when $H=0$ because Dirichlet conditions on $u$ imply Dirichlet conditions on $\Delta u$, simply by virtue of the eigenvalue equation, and the orthogonality constraint eqn (\ref{orthogconstraint}) is trivially satisfied. As a consequence, the eigenvalues of $\mathcal{L}$ take the form $\lambda_m=\mu^{(m)}(\mu^{(m)}-\mu^{(n)})/(2\kappa)$ for $m=1,2,\hdots$ for the $n$th mode catenoid. 
From this expression, we can deduce for the thick catenoid, whose allowable tensions $\mu^{(m)}$ are all negative, that the $n=1$ mode is (marginally) stable while the higher order modes have at least one negative eigenvalue $\lambda_{n-1}$ and are hence  unstable. For the thin catenoid, which has $\mu^{(1)}>0$, the first and second modes are both (marginally) stable, while the higher order modes are unstable.
This situation stands in stark contrast to the case of the soap film, where thin catenoids are always unstable and thick catenoids are always stable with respect to the area functional.


When $H\neq 0$, our numerical results indicate that the minimal eigenvalue tends to decrease as $n$ is increased, so that higher order surfaces tend to be unstable. For the three values of $\bar{A}$ explored in this paper, no shape with $n>4$ was found to be stable.
For $\bar{A} = 1$, only the $n=1$ branches are stable, and this is only when the extension $h/a \gtrsim 0.1$. (Fig.~\ref{fig:Fvh_smallA}). For $\bar{A} > 1$, thin branches can contain stable shapes for sufficiently small $n$, as Figs.~\ref{fig:Fvh_medA} and ~\ref{fig:Fvh_largeA} show.
Notably, the tethers that appear when $\bar{A} > 1$ are stable, in agreement with previous work~\cite{powers_huber_goldstein2002}.

\subsubsection{Case of nonzero Gaussian curvature modulus}

For the reasons discussed above, the minimal eigenvalue diverges when $\bar\kappa \neq 0$ and $\bar{A} < \bar{A}_\mathrm{max}$. The direction of divergence of the minimal eigenvalue is the same as the that of the force, as illustrated in Fig.~\ref{fig:kbarshapes}c. This shows that at extensions near $h=h^*$, nonzero $\bar\kappa$ has a stabilizing effect on one branch and a destabilizing effect on the other. In particular, a nonzero $\bar\kappa$ is necessary to stabilize higher order catenoid-like surfaces. Changing the sign of $\bar\kappa$ changes the direction of divergence. 

When $\bar{A} > \bar{A}_\mathrm{max}$, the divergence at the maximal extension is not present, as noted before. Values of $|\bar\kappa|<\kappa$ are observed to have a negligible effect on the stability of these surfaces. The tether remains a stable solution.

\section{Conclusions}
\label{discuss_and_conclude}

\begin{figure*}[t]
\centering
\includegraphics[scale=0.18]{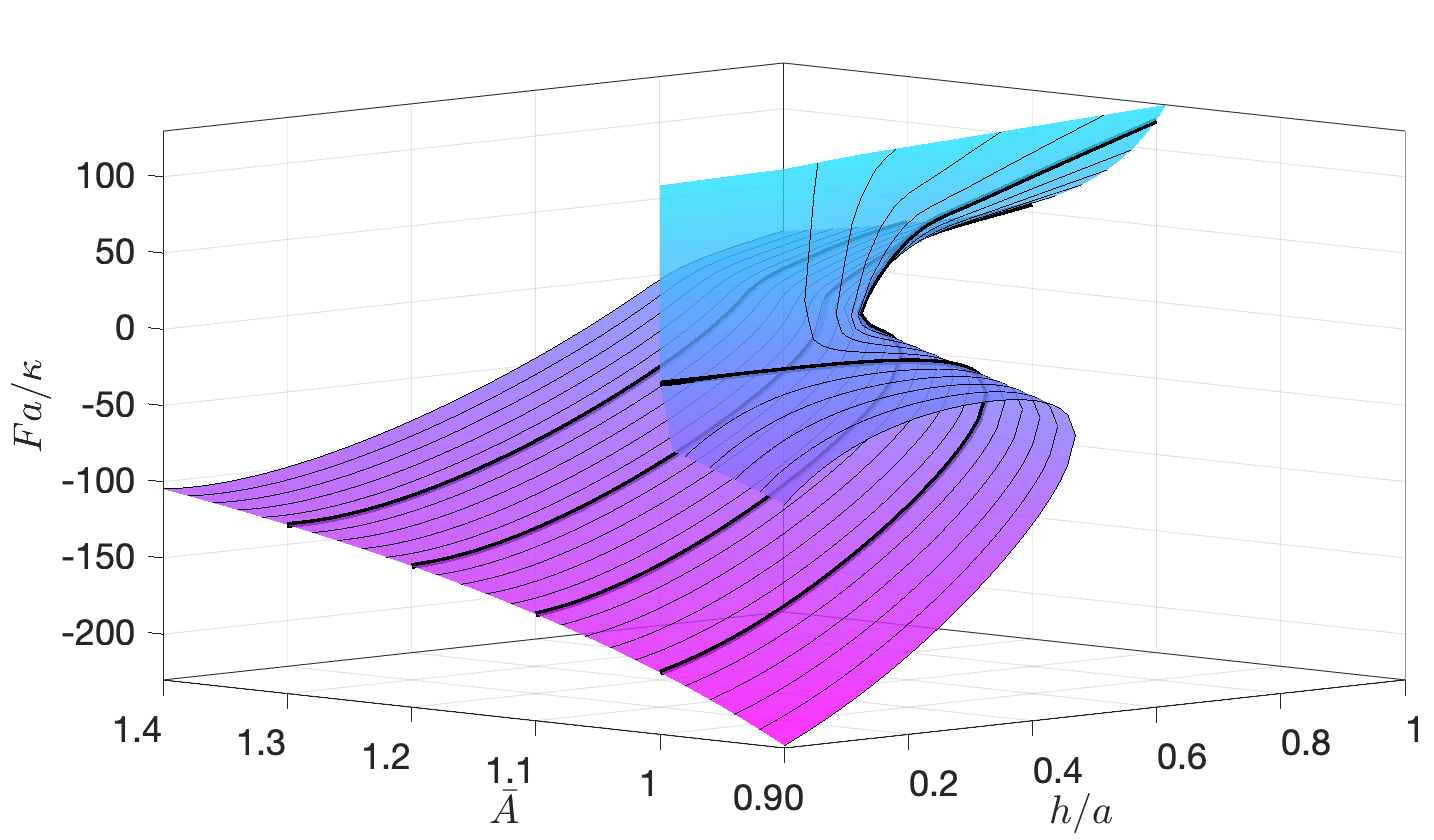}\hfill
\includegraphics[scale=0.18]{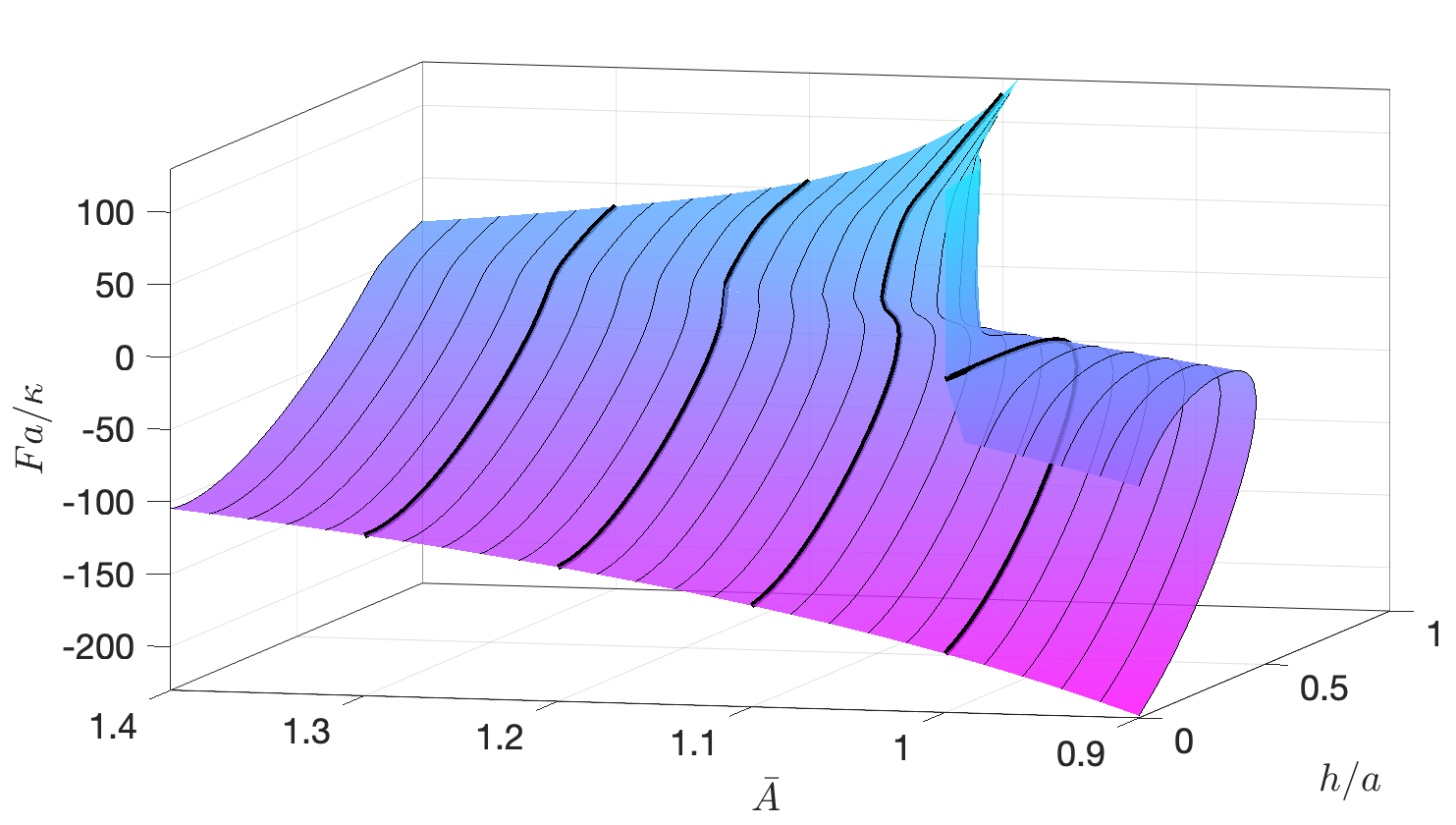}
\caption{Two views of the manifold of lowest mode equilibrium solutions to the axisymmetric membrane pulling problem with zero Gaussian curvature modulus in $(h/a,\bar{A},Fa/\kappa)$-space, showing a cusp-type catastrophe at $\bar{A} = \bar{A}_{\mathrm{max}}=1.1997$, where the catenoid ceases to exist. The manifold runs off to infinite force at $\bar{A} = 1^+$ and also at $h = 0$ (not shown), where the thin catenoid has vanishingly small neck radius. Black curves are level sets of constant $\bar{A}$ with the thick black lines spaced by $0.1$ from $\bar{A}=1$ to $1.3$; the lines at $\bar{A}=1$, $1.1$, and $1.3$ correspond to the $n=1$ curves of Figs.~\ref{fig:Fvh_smallA}, \ref{fig:Fvh_medA}, and \ref{fig:Fvh_largeA}, respectively. Color indicates dimensionless force.}
\label{fig:eqmanifold}
\end{figure*}

We have generalized the time-honored soap film Plateau problem to the stretching of a fixed-area fluid membrane suspended between two symmetric rings. In so doing, we have unified various classic shapes such as catenoids, thin tethers, and Willmore tori, as well as new buckled oscillatory shapes, as different limits of a single system, with area serving as a bifurcation parameter. Fig.~\ref{fig:eqmanifold} summarizes how the force vs. extension curve changes with area. Since we enforce fixed area, the tension must be determined, and the membrane shape  equation eqn (\ref{shapeEq}) becomes a nonlinear eigenvalue problem, generally yielding infinitely many solution branches for a given extension. 
Particular attention was paid to the catenoid, which always appears at a local (and global, if $\bar{A}<1$) extremum of extension.  By formulating the catenoid-pulling problem as a perturbation problem, we calculated its permissible values of force (in the zero Gaussian curvature modulus case) and its singular behavior (in the nonzero Gaussian curvature modulus case). The tension and stability of catenoidal membranes were also shown to be directly connected to the stability of catenoidal soap films, by means of the Jacobi operator $\mathcal{J}$ for minimal surfaces.

Although the model described in this paper was initially conceived for axisymmetric  colloidal membranes, colloidal membranes have more degrees of freedom that we do not account for here. For example, a more general model could remove the fixed ring assumption and balance forces at the boundary rings, perhaps with an edge bending stiffness and a line tension. A more ambitious model might build in liquid crystalline rod-rod interactions.



Future work could also include spontaneous curvature, where Delaunay surfaces could appear as possible solutions. There are also other asymmetric or self-intersecting solutions we didn't cover in depth in this paper. Analysis of these shapes may strengthen the analogy between these surfaces and the classical elastica. It would also be interesting to study the forces associated with membrane transitions analogous to the transition between a helicoid and a catenoid seen in a soap film~\cite{BoudaoudPatricioBenAmar1999}, or the topological transition 
 transformation from M\"obius strip to two-sided soap film~\cite{GoldsteinMoffattPesciRicca2010}.

\section*{Conflicts of interest}
There are no conflicts to declare.

\section*{Acknowledgments}
This work was supported in part by the National Science Foundation through Grants No. MRSEC-1420382, No. CMMI-1634552, and CMMI-2020098. Some of this work was completed while TRP was a participant in the research program on ``Growth, Form, and Self-organization" at the Isaac Newton Institute for Mathematical Sciences in Cambridge, UK, funded by National Science Foundation Grant No. PHY-1708061. We are grateful to Anthony Dinsmore, Zvonimir Dogic, Benjamin Friedrich, Raymond Goldstein, Jemal Guven, James Hanna, and Megan Kerr for helpful discussion.

\appendix

\section{Appendices}

\subsection{Geometrical formulas}
 \label{dgappx}

Here we define all the geometric quantities we use for completeness, and especially to make our sign conventions clear. We begin by writing the general geometric formulas, and then specialize to the coordinates of Fig.~\ref{fig:setup}. A surface is given by the parametrization $\mathbf{X}(\xi^1,\xi^2)$, where $\xi^1$ and $\xi^2$ are coordinates. The first fundamental form is given by 
\begin{equation}
  I=\mathrm{d}\mathbf{X}\cdot\mathrm{d}\mathbf{X}=g_{ij}\mathrm{d}\xi^i\mathrm{d}\xi^j,
  \end{equation} where $g_{ij}=\mathbf{t}_i\cdot\mathbf{t}_j$ is the metric tensor. 
  Thus, the area element is $\mathrm{d}A=\sqrt{g}\mathrm{d}^2\xi$, where $g$ denotes the determinant of the metric tensor. Also, the outward normal to the surface is given by $\hat{\mathbf n}=\partial\mathbf{X}/\partial\xi^1\times\partial\mathbf{X}/\partial\xi^2/\sqrt{g}$,
  and the second fundamental form is given by 
  \begin{equation}
  II=-\mathrm{d}\mathbf{n}\cdot\mathrm{d}\mathbf{X}=K_{ij}\mathrm{d}\xi^i\mathrm{d}\xi^j,
  \end{equation}
  where $K_{ij}$ is the curvature tensor. As usual we raise indices with $g^{ij}$, the inverse of the metric tensor; for example, $K^i_j=g^{ik}K_{kj}$. The mean curvature is $H=K^i_i/2$ and Gaussian curvature is $K=\det K^i_j$.
The Laplacian operator is defined to be
 \begin{equation}
 \Delta = \frac{1}{\sqrt{g}} \frac{\partial }{\partial \xi^i}  \sqrt{g} g^{ij} \frac{\partial }{\partial \xi^j}. 
 \end{equation}
 As described in discussion of eqn~(\ref{notorque}), the normal curvature $k_\mathrm{n}$ of a boundary curve $C$ of the surface is given by $k_\mathrm{n}=\mathbf{n}_C\cdot\mathrm{d}\hat{\mathbf{T}}/\mathrm{d}\ell$, where $\mathbf{n}_C$ is the surface normal to the edge, $\hat{\mathbf T}$ is the unit tangent vector of the edge, and the direction of increasing arclength $\ell$ along the edge is such that the surface lies to the left of edge as it is traversed. 

For an axisymmetric surface we use the coordinates $s$ and $\phi$, where $s$ is arclength of the meridian, and $\phi$ is the azimuthal angle (Fig.~\ref{fig:setup}). The position in Cartesian components of a point with coordinates $(s,\phi)$ is $\mathbf{X}=(r(s)\cos\phi, r(s)\sin\phi,z(s))$, which leads to the tangent vectors 
\begin{eqnarray}
\mathbf{t}_1&=&(\cos\psi\cos\phi,\cos\psi\sin\phi,-\sin\psi)\\
 \mathbf{t}_2&=&(-r\sin\phi,r\cos\phi,0).
 \end{eqnarray}
  Thus, the first and second fundamental forms are given by 
  \begin{eqnarray}
  I&=&\mathrm{d}s^2+r^2\mathrm{d}\phi^2\\
  II&=&-\frac{\mathrm{d}\psi}{\mathrm{d}s}\mathrm{d}s^2-r\sin(\psi)\mathrm{d}\phi^2.
  \end{eqnarray}   The area element is $\mathrm{d}A=r\mathrm{d}s\mathrm{d}\phi$, the outward normal to the surface is $\hat{\mathbf n}=(\sin\psi\cos\phi,\sin\psi\sin\phi,\cos\psi)$, and the mean and Gaussian curvature are
  \begin{eqnarray}
 H&=& -\frac{1}{2}\left(\frac{\mathrm{d}\psi}{\mathrm{d}s}+\frac{\sin\psi}{r}\right)\label{Hitopsirs}\\
 K&=&\frac{\sin(\psi)}{r}\frac{\mathrm{d}\psi}{\mathrm{d}s}.\label{eqnKformula}
  \end{eqnarray}
Finally, the Laplacian operator for an axisymmetric surface is 
 \begin{equation}
 \Delta  = \frac{1}{r} \frac{d}{ds} r \frac{d}{ds}. \label{eqnnablaform} 
 \end{equation}

 \subsection{The Noether invariant and axial force}\label{NoethInvt}

In this section we derive an expression for the Noether invariant associated with the energy eqn~(\ref{EEq}) under translation along the axis of symmetry and show that this invariant is in fact the axial force. For the purposes of this derivation it is convenient to write
the energy as $E=\int\mathfrak{E}\mathrm{d}z$, where $\mathfrak{E}(r, r_z,r_{zz})\mathrm{d}z=\mathcal{E}(r,r_s,r_{ss})\mathrm{d}s$, with the $z$-axis the axis of symmetry, rather than using the parameterization of Sec.~\ref{parameterization}.
Since the energy density has no explicit dependence on the variable $z$, $\partial\mathfrak{E}/\partial z=0$, and we may write
\begin{equation}
\frac{\mathrm{d}\mathfrak{E}}{\mathrm{d}z}=\frac{\partial\mathfrak{E}}{\partial r}r_z+\frac{\partial\mathfrak{E}}{\partial r_z}r_{zz}+\frac{\partial\mathfrak{E}}{\partial r_{zz}}r_{zzz}.\label{noddz}
\end{equation}
On the other hand, the Euler-Lagrange equation for $r(z)$ is
\begin{equation}
\frac{\partial\mathfrak{E}}{\partial r}-\frac{\mathrm{d}}{\mathrm{d}z}\frac{\partial \mathfrak{E}}{\partial r_z}+\frac{\mathrm{d}^2}{\mathrm{d}z^2}\frac{\partial \mathfrak{E}}{\partial r_{zz}}=0.\label{EL}
\end{equation}
Using eqn~(\ref{EL}) to eliminate $\partial\mathfrak{E}/\partial r$ from eqn~(\ref{noddz}), and using the Leibniz rule to rearrange some derivatives, we find that
\begin{equation}
\mathfrak{E}-r_z\frac{\partial \mathfrak{E}}{\partial r_z}-r_{zz}\frac{\partial \mathfrak{E}}{\partial r_{zz}}+r_z\frac{\mathrm d}{{\mathrm d}z}\left(\frac{\partial \mathfrak{E}}{\partial r_{zz}}\right)\equiv F\label{bigNoether}
\end{equation}
is constant~\cite{Logan1977}.

This conserved quantity can be shown to be the axial force by using the principle of virtual work:
\begin{equation}
    \delta E-F_1\delta z_1-\int\mathrm{d}\phi r_1 m\delta\theta_1=0,
\end{equation}
where $F_1$ is the external force, $z_1$ is the $z$-position of the ring subject to the virtual displacement $\delta z_1$, $r_1$ is the radius of the membrane  at $z_1$, $m$ is the external bending moment per unit length at $z_1$, and $\delta \theta_1$ the virtual change in the angle $\theta_1$ defined by $\tan\theta_1=r_z(z_1)$. First consider the variation $\delta E$ due to the change in radius $\zeta(z)$ and change in ring position (the ring at $-z_1=-h/2$ remains fixed):
\begin{equation}
  \delta E=\int_{-z_1}^{z_1+\delta z_1}\left[{\mathfrak E}(r+\zeta,r_z+\zeta_z,r_{zz}+\zeta_{zz})-{\mathfrak E}(r,r_z,r_{zz})\right].
\end{equation}
Expanding to first order $\zeta$ and $\delta z_1$, and integrating by parts as usual, we find
\begin{eqnarray}
    \frac{\delta E}{2\pi}&=&\int_{-z_1}^{z_1}\frac{\delta E}{\delta r}\mathrm{d}z\nonumber\\
    &+&\left[\zeta\left(\frac{\partial \mathfrak{E}}{\partial r_z}-\frac{\mathrm{d}}{\mathrm{d}z}\frac{\partial\mathfrak{E}}{\partial r_{zz}}\right)+\zeta_z\frac{\partial \mathfrak{E}}{\partial r_{zz}}+{\mathfrak E}\delta z_1\right]_{z_1},
\end{eqnarray}
where
\begin{equation}
    \frac{\delta E}{\delta r}=\frac{\partial \mathfrak{E}}{\partial r}-\frac{\mathrm{d}}{\mathrm{d}z}\frac{\partial \mathfrak{E}}{\partial r_z}+\frac{\mathrm{d}^2}{\mathrm{d}z^2}\frac{\partial \mathfrak{E}}{\partial r_{zz}}.
\end{equation}
To make progress, we must relate $\zeta(z_1)$ and $\zeta_z(z_1)$ to $\delta z_1$ and $\delta\theta_1$ using~\cite{GelfandFomin1963}
\begin{eqnarray}
 r(z_1)+\delta r_1&=&r(z_1+\delta z_1)+\zeta(z_1+\delta z_1)\\
 \tan(\theta_1+\delta\theta_1)&=&r_z(z_1+\delta z_1)+\zeta_z(z_1+\delta z_1),
\end{eqnarray}
or, working to first order in the small quantities,
\begin{eqnarray}
 \zeta(z_1)&=&\delta r_1-\delta z_1 r_z(z_1)\\
 \zeta_z(z_1)&=&\sec^2\theta_1\delta\theta_1-\delta z_1 r_{zz}(z_1).
\end{eqnarray}
Using these formulas for $\zeta$ and $\zeta_z$ at the displaced end, we find
\begin{eqnarray}
 &&(\delta E-F_1\delta z_1)-2\pi r_1 m\delta\theta_1=2\pi\int_0^{z_1}\frac{\delta E}{\delta r}\mathrm{d}z\nonumber\\
 &+&\left[\mathfrak{E}-r_z\frac{\partial\mathfrak{E}}{\partial r_z}+r_z\frac{\mathrm{d}}{\mathrm{d}z}\frac{\partial \mathfrak{E}}{\partial r_{zz}}-r_{zz}\frac{\partial \mathfrak{E}}{\partial r_{zz}}-{F_1}\right]_{z_1}\delta z_1\label{bigvar}\\
 &+&\left[\sec^2\theta_1\frac{\partial\mathfrak{E}}{\partial r_{zz}}-2\pi r_1 m\right]_{z_1}\delta\theta_1+
 \left[\frac{\partial \mathfrak{E}}{\partial r_z}-\frac{\mathrm{d}}{\mathrm{d}z}\frac{\partial\mathfrak{E}}{\partial r_{zz}}\right]_{z_1}\delta r_1.\nonumber
\end{eqnarray}
  Since $\delta E-F_1\delta z_1-(2\pi)r_1 m\delta\theta_1=0$, we conclude that the axial force is the Noether invariant found in eqn~(\ref{bigNoether}), $F=F_1$. Note that $\delta r_1=0$ for our rigid rings, but we could use the coefficient of $\delta r_1$ in eqn~(\ref{bigvar}) to find the radial force per unit length of the membrane on the ring. Also, the bending moment per unit length is given by $m=\partial\mathfrak{E}/\partial r_{zz}/( 2\pi r_1\cos^2\theta_1)$. Using the following formulas for axisymmetric shapes, which arise from $\mathbf{X}=(r(z)\cos\phi,r(z)\sin(\phi),z)$ and the formulas of the preceding section,
 \begin{eqnarray}
  \sqrt{g}&=&r\sqrt{1+r_z^2}\nonumber\\
  2H&=&\frac{r_{zz}}{(1+r_z^2)^{3/2}}-\frac{1}{r(1+r_z^2)^{1/2}}\label{Hitorz}\\
  K&=&-\frac{r_{zz}}{r(1+r_z^2)^2}\\
  k_n&=&-\frac{1}{r_1(z_1)\sqrt{1+r_z(z_1)^2}},
 \end{eqnarray}
 one may verify that our formula for $m$ gives the expected total edge bending moment~\cite{TuOu-Yang2004} $M=2\pi m r_1=2\kappa H+\bar{\kappa} k_n$.

In the case of a soap film between two rings with centers on the $z$-axis, the axial force $F$ needed to hold the two rings apart can be found from eqn~(\ref{bigNoether}) with the result
\begin{equation}
\mathfrak{E}-r_z\frac{\partial \mathfrak{E}}{\partial r_z}=2\pi\mu\frac{r}{\sqrt{1+r_z^2}}= F,\label{soapfilmF}
\end{equation}
where we have used the soap film energy density $\mathfrak{E}=2\pi\mu r\sqrt{1+r_z^2}$.  When $r=b$, where $b$ is the neck radius of the catenoid, i.e.,   the smallest radius of the catenoid, then  $r_z=0$ and $F=2\pi\mu b$ as expected.

For the general energy eqn~(\ref{EEq}) we have,
\begin{eqnarray}
\mathfrak{E}&=&2\pi\mu r\sqrt{1+r_z^2}\nonumber\\
&+&2\pi\frac{\kappa}{2}r\sqrt{1+r_z^2}\left[\frac{r_{zz}}{(1+r_z^2)^{3/2}}-\frac{1}{r(1+r_z^2)^{1/2}}\right]^2,
\end{eqnarray}
where we have not written the Gaussian curvature term because direct calculation shows it makes make no contribution to $F$.
The full expression for $F$ in this case of nonzero bending modulus $\kappa$
is too complicated to quote,
but if we choose the origin of $z$ to coincide with the neck, where $r_z=0$, then 
\begin{equation}
F=2\pi b\mu + 2\pi\kappa\frac{1-b^2 r_{zz}^2(0)}{2b}.
\label{forceEq}
\end{equation}
Eqn~(\ref{forceEq}) can be seen to be equivalent to eqn~(\ref{forceetaeqn}) by evaluating $\mathcal{H}$ at the neck and using the fact that $\mathcal{H} = 0$.
For a free-floating shape, $F=0$, and we find a relation between the membrane tension, membrane bending stiffness, and longitudinal curvature at $z=0$:
\begin{equation}
\mu=\frac{\kappa}{2}\left[ r_{zz}^2(0)-\frac{1}{b^2}\right].\label{muEq}
\end{equation}

The sign of the Gaussian curvature modulus only affects the bending moment boundary condition $2\kappa H+\bar{\kappa}k_\mathrm{n}=0$. Positive $\bar{\kappa}$ makes the mean curvature at the edge negative, and negative $\bar{\kappa}$ makes the mean curvature at the edge positive.  Unduloids and spheres have negative mean curvature in our convention, whereas nodoids have positive mean curvature (e.g. see ref~\cite{BenditoBowickMedina2014}, whose sign convention is opposite to ours).

 \bibliography{newrefs}
 \bibliographystyle{rsc}

\end{document}